\documentclass[12pt]{article}
\usepackage{amsfonts,amssymb,amsbsy,latexsym,amsmath,tabulary,graphicx,times,xcolor}
\usepackage[T1]{fontenc}
\usepackage{url,multirow,morefloats,floatflt,cancel,tfrupee}
\usepackage{colortbl}
\usepackage{xcolor}
\usepackage{pifont}
\usepackage[nointegrals]{wasysym}
\usepackage{gensymb}
\usepackage{setspace}
\usepackage{tablefootnote}
\usepackage[paperheight=11in,paperwidth=8.5in,margin=1in,headsep=.5cm,top=2.5cm]{geometry}

\makeatletter

\def\author#1{\gdef\@author{\hskip-\dimexpr(\tabcolsep)\hskip1pt\parbox{\dimexpr\textwidth-12pt}{\center \fontsize{12}{14}\selectfont#1}}}
\def\title#1{\gdef\@title{\center\vspace{-1cm}\ifx\@articleType\@empty\else\@articleType\\\fi\selectfont#1}}
\let\@articleType\@empty \def\articletype#1{\gdef\@articleType{{\textbf\normalfont#1}}}
\makeatother

\date{}

\usepackage{lineno}

\usepackage[authoryear,sort&compress,comma]{natbib}
\setcitestyle{aysep={}}

\usepackage{float}

\begin{document}

\title{Detection and characterization of microseismic events from fiber-optic DAS data using deep learning}

\author{Fantine Huot\textsuperscript{1}, Ariel Lellouch\textsuperscript{1,3}, Paige Given\textsuperscript{1}, Bin Luo\textsuperscript{1}, Robert G. Clapp\textsuperscript{1}, \\
Tamas Nemeth\textsuperscript{2}, Kurt T. Nihei\textsuperscript{2}, and Biondo L. Biondi\textsuperscript{1} \\~\\
\textsuperscript{1} Stanford University \\
397 Panama Mall \\
Stanford, CA 94025 \\
USA, \\
\textsuperscript{2} Chevron Technical Center \\
1500 Louisiana St \\
Houston, TX 77002 \\
USA, \\
\textsuperscript{3} Tel Aviv University \\
Tel Aviv 6997801 \\
Israel 
\\~\\
{Email: fantine@sep.stanford.edu (Fantine Huot)}
}

\maketitle

\newpage

\begin{abstract}
Microseismic analysis is a valuable tool for fracture characterization in the earth's subsurface. As distributed acoustic sensing  (DAS) fibers are deployed at depth inside wells, they hold vast potential for high-resolution microseismic analysis. However, the accurate detection of microseismic signals in continuous DAS data is challenging and time-consuming. We design, train, and deploy a deep learning model to detect microseismic events in DAS data automatically. We create a curated dataset of nearly 7,000 manually-selected events and an equal number of background noise examples. We optimize the deep learning model's network architecture together with its training hyperparameters by Bayesian optimization. The trained model achieves an accuracy of 98.6\% on our benchmark dataset and even detects low-amplitude events missed during manual labeling. Our methodology detects more than 100,000 events allowing the reconstruction of  spatio-temporal fracture development far more accurately and efficiently than would have been feasible by traditional methods.                       
\end{abstract}

\newpage
\section*{Introduction}

As humans interact with the Earth's natural resources, our impact on climate and energy reserves is becoming increasingly apparent, prompting the development of renewable energies and carbon sequestration. A growing proportion of these technological advances will require subsurface monitoring techniques. Geothermal energy, for instance, has massive potential for low-carbon power generation -- the amount of heat within 10 km of the surface is estimated to contain 50,000 times more energy than all oil and gas resources worldwide (\citet{shere2013renewable}). In parallel, in order to limit global warming to well below 2\degree C, estimations show that carbon capture and storage (CCS) efforts have to capture 800 to 1,200 gigatons of CO\textsubscript{2} by the end of the century (\citet{lal2004soil,metz2005ipcc,iea2019world}). 
Such large-scale CCS would involve the creation of thousands to tens of thousands of geological sequestration sites (\citet{lal2004soil,metz2005ipcc,iea2019world}). The safety of these geothermal and CCS facilities requires monitoring technologies to assess the subsurface's structural health (\citet{oldenburg2009certification,wilson2007carbon}).  Fiber-optic acquisition combined with microseismic analysis could be a valuable tool for such subsurface monitoring tasks.

Distributed acoustic sensing (DAS) is a technology that uses fiber-optic cables to record the seismic strain, or strain-rate, along the direction of the cable. DAS has been increasingly adopted for industrial applications such as the monitoring of CO\textsubscript{2} sequestration (\citet{daley2013field}), hydraulic stimulation (\citet{bakku2015fracture}), and geothermal sites (\citet{lellouch2020comparison,lellouch2021low}). In recent years, DAS has been used for microseismic monitoring (\citet{Karrenbach2019,Baird2020,Stork2020}) because of its extensive spatial coverage and high spatio-temporal resolution. DAS cables can be deployed at depth in existing wells at low cost, close to where microseismic events originate. Their dense spatial coverage over long distances has great potential for recording and characterizing many more microseismic events than sparse sensors located at the surface (\citet{lellouch2020comparison}) and can even be used to image subsurface fractures (\citet{lellouch2019observations}). 

Microseismic analysis is one of the primary tools for fracture characterization. It is particularly valuable if it can be conducted in time to influence engineering and production decisions. However, DAS measurements produce large data volumes (over 10$\,$TB/day for modern DAS acquisitions), so new automated processing workflows must be developed to utilize their potential for microseismic monitoring. In addition, DAS data are generally unsuitable for thresholding-based detection methods because individual DAS channels are noisier than conventional borehole instruments (\citet{lellouch2020comparison}). Other detection methods such as template matching take advantage of the spatio-temporal recording, but microseismic events can vary in location, focal mechanism, and wave propagation characteristics, resulting in significantly varying recorded signals. 

Recent studies demonstrate that machine learning models have great potential for waveform-based event detection on continuous DAS data, with applications ranging from earthquake (\citet{huot2018jump,huot2020detecting}) to vehicle detection (\citet{huot2017automatic,martin2018seismic,huot2018automated,huot2018machine}). For microseismic event detection, \cite{Stork2020} generate synthetic microseismic events using an anisotropic forward modeling operator. They use these synthetic examples to train YOLOv3, a state-of-the-art neural network for image object detection. In another study, \cite{Binder2019} use a homogenous elastic medium for the forward modeling and train a network of their design for microseismic detection. Both approaches supply background noise examples to the neural network by selecting noise from inactive periods of the continuous DAS records. Both methods were found relatively successful but underperformed when compared to human labeling (\citet{Stork2020}) and to a surface-based microseismic catalog (\citet{Binder2020}). These studies were also limited to relatively small datasets containing hundreds of events at most.

In this study, we train a convolutional neural network (CNN) for microseismic monitoring over continuous DAS data. We illustrate our methodology on field data acquired in a horizontal well during hydraulic stimulation, a technique commonly used in low-permeability rocks like tight sandstone or shale to increase gas flow. A similar approach is used to enhance permeability in underground geothermal reservoirs. Man-made reservoirs called  Enhanced Geothermal Systems (EGS) can generate geothermal electrcity without the need for natural convective hydrothermal resources.

We create a unique curated catalog of 6,786 manually-selected microseismic events and choose an equal number of random noise examples. Our resulting dataset contains thus 13,572 field data examples with events of varying locations, amplitudes, and focal mechanisms, making it suitable for machine learning applications without the need to generate additional synthetic data. This is valuable because certain wave propagation characteristics, such as dispersive guided waves present in this dataset, are difficult to replicate synthetically. 

We design the neural network architecture in a modular fashion using heuristics from state-of-the-art deep learning models while controlling the details of the architecture with several hyperparameters.  We do so by creating modular versions of known CNN classifiers, such as VGG (\cite{simonyan2015deep}) and ResNet (\cite{he2015deep}). We create flexibility in these network architectures by parameterizing many elements such as network depth, width, and types of layers. Tuning the hyperparameters allows us to tailor these CNNs to the task at hand. This approach strikes a compromise between exploiting architectures known to perform well and exploring new models. 

Hyperparameters heavily influence the behavior of the learned model and must be carefully selected to ensure the network's performance (\citet{goodfellow2016deep}). High accuracy is especially relevant in the context of detection algorithms meant to scale over continuous data, where slight differences in detection accuracy quickly translate to thousands of false detections. We frame the hyperparameter tuning problem as Bayesian optimization and search for the best network performance as a function of the hyperparameters. Bayesian optimization has been increasingly adapted for hyperparameter search and is now part of some open-source machine learning frameworks (\cite{feurer2019auto,gpyopt2016}). These software packages cover the hyperparameter search itself but not the network design, and as such, they do not offer the type of modular architecture exploration we present here. It helps us quickly identify promising deep learning models, allowing us to incorporate them into our geophysical processing workflow. We share our hyperparameter tuning code for users to adapt to their use
cases\footnote{https://github.com/fantine/microseismic-detection-ml}.

We first present the dataset and the signal processing workflow. We then describe the neural network architecture and hyperparameter tuning. We demonstrate that this deep learning model accurately detects microseismic signals and even retrieves small events missed during manual labeling. We scale the workflow over continuous data, detecting more than 100,000 new microseismic events, allowing us to create a detailed mapping of the fracture development that would not have been feasible using traditional methods on a single fiber.

\section{Data acquisition and processing}

\subsection{Data acquisition}

\begin{figure}
    \begin{center}
    \includegraphics[width=0.6\linewidth,trim={0cm 1.5cm 0cm 2cm},clip]{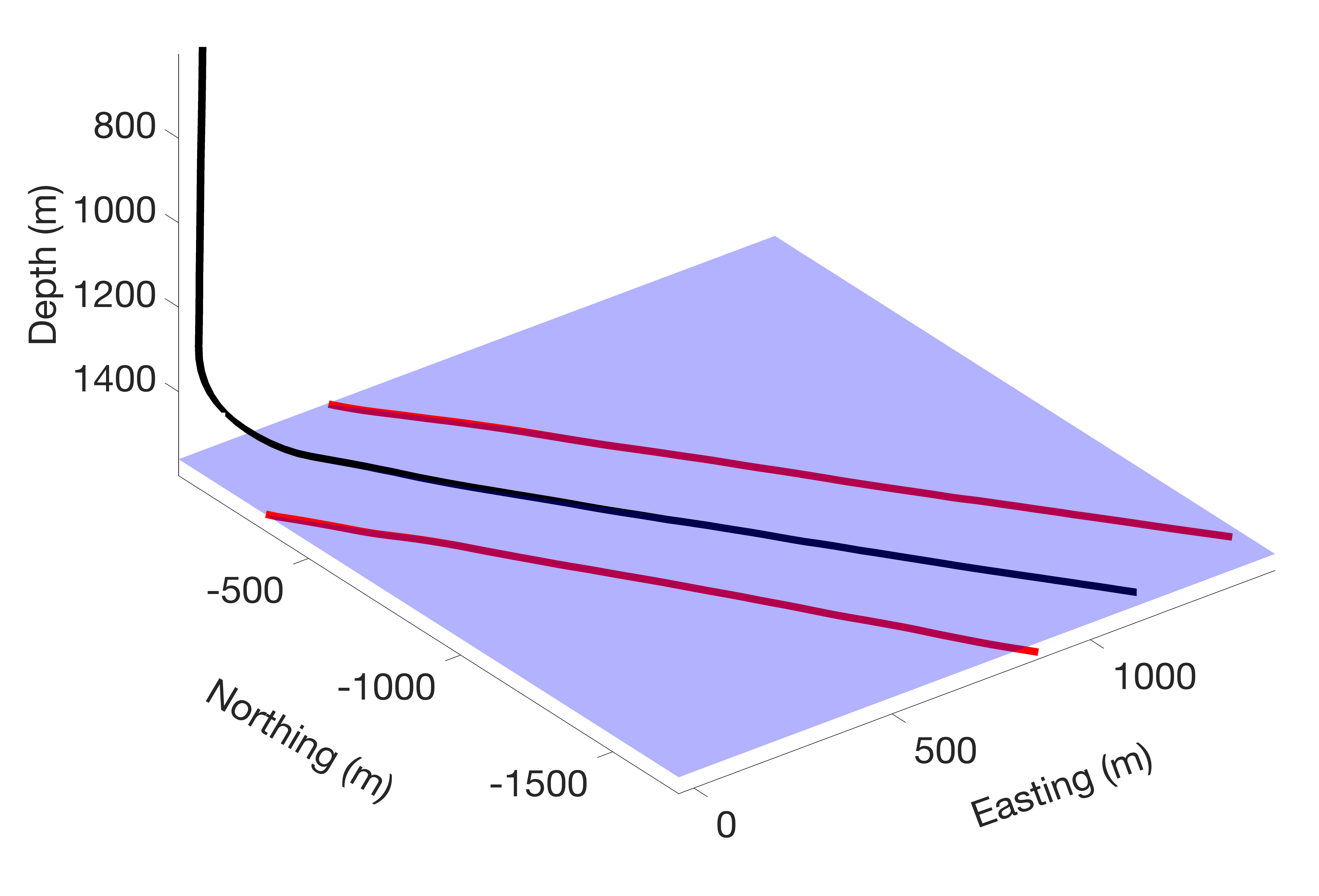}
    \end{center}
    \caption{Well instrumented with fiber (black) and horizontal sections of the offset wells (red). 
    The offset wells were stimulated alternatively using a zipper-fracturing schedule, generating the microseismic activity recorded by the fiber.
    }
    \label{fig:acquisition}
\end{figure}

The data were acquired by a DAS fiber deployed in a deviated well drilled into an unconventional shale reservoir. The target reservoir is 15$\,$m thick with significantly lower seismic velocities and density than its surroundings. The DAS array continuously records stimulation activity from two offset wells positioned in the reservoir and located approximately 270$\,$m away on the North-East and South-West sides of the DAS well (Figure~\ref{fig:acquisition}). Their horizontal sections, almost 2$\,$km long, are inside the reservoir and, up to a few meters, at the same depth. The acquisition setup and reservoir properties are summarized in \cite{lellouch2019observations,Lellouch2020b}. 

The DAS data were recorded in 2015 with an OTDR interrogator (iDAS\texttrademark \hspace{0.05cm} from Silixa) measuring strain rate at a 1$\,$m channel spacing and 10$\,$m gauge length (distance over which the strain measurement is acquired), over 4,000 channels at 2,000 samples per second, representing 1.5$\,$TB of data per day of recording. The offset wells were perforated and stimulated alternatively using a zipper-fracturing schedule, generating the microseismic activity recorded by the DAS fiber. Stimulation shots in both offset wells are contiguous and located approximately 12$\,$m apart.  Data were recorded continuously for about two weeks and covered 58 stimulation stages, with 31 and 27 stages in the two offsets wells, respectively.

\subsection{Data labeling}
\label{sec:data_labeling}

\begin{figure}
    \begin{center}
    \includegraphics[width=0.42\linewidth]{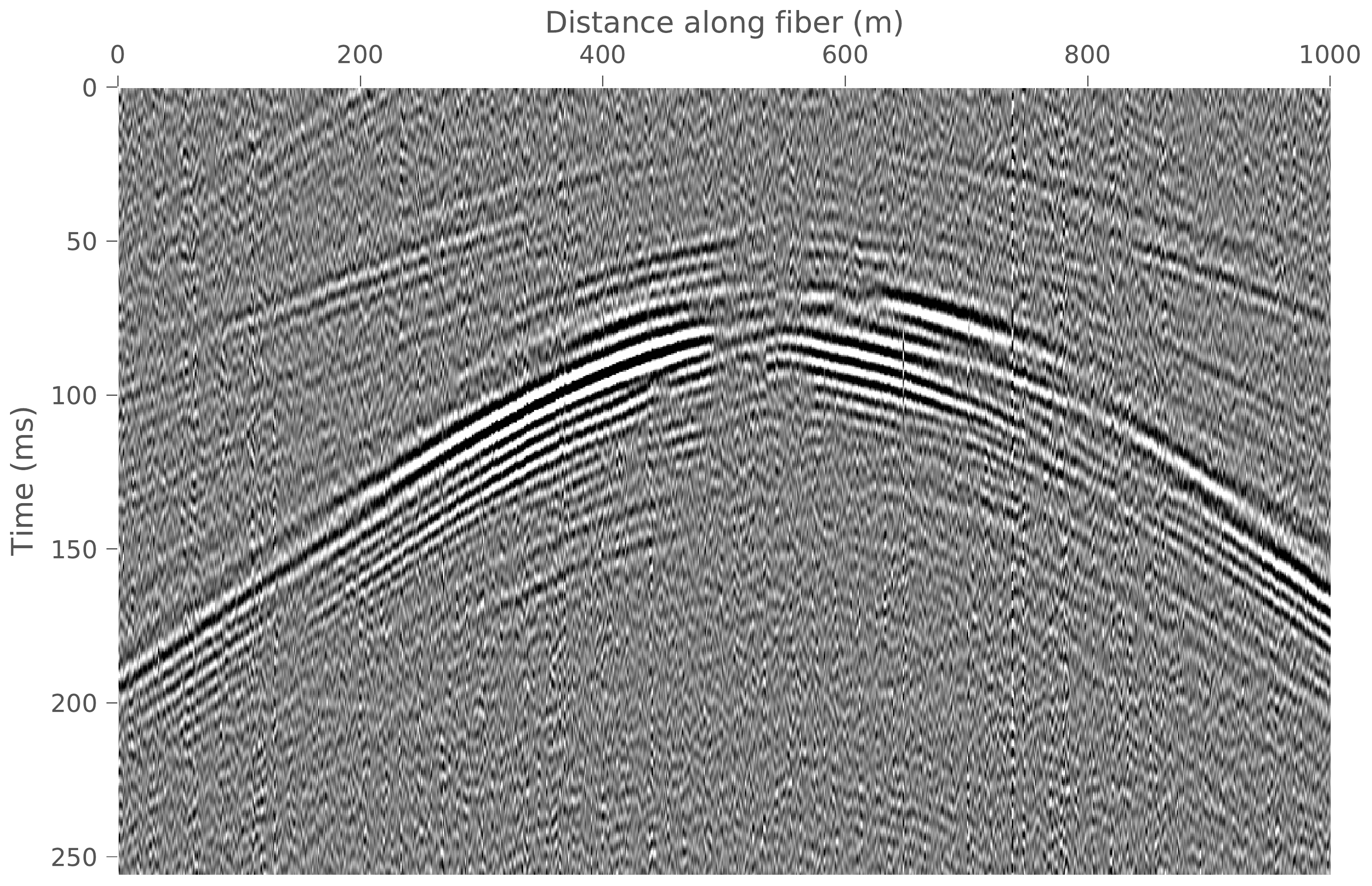}
    \includegraphics[width=0.42\linewidth]{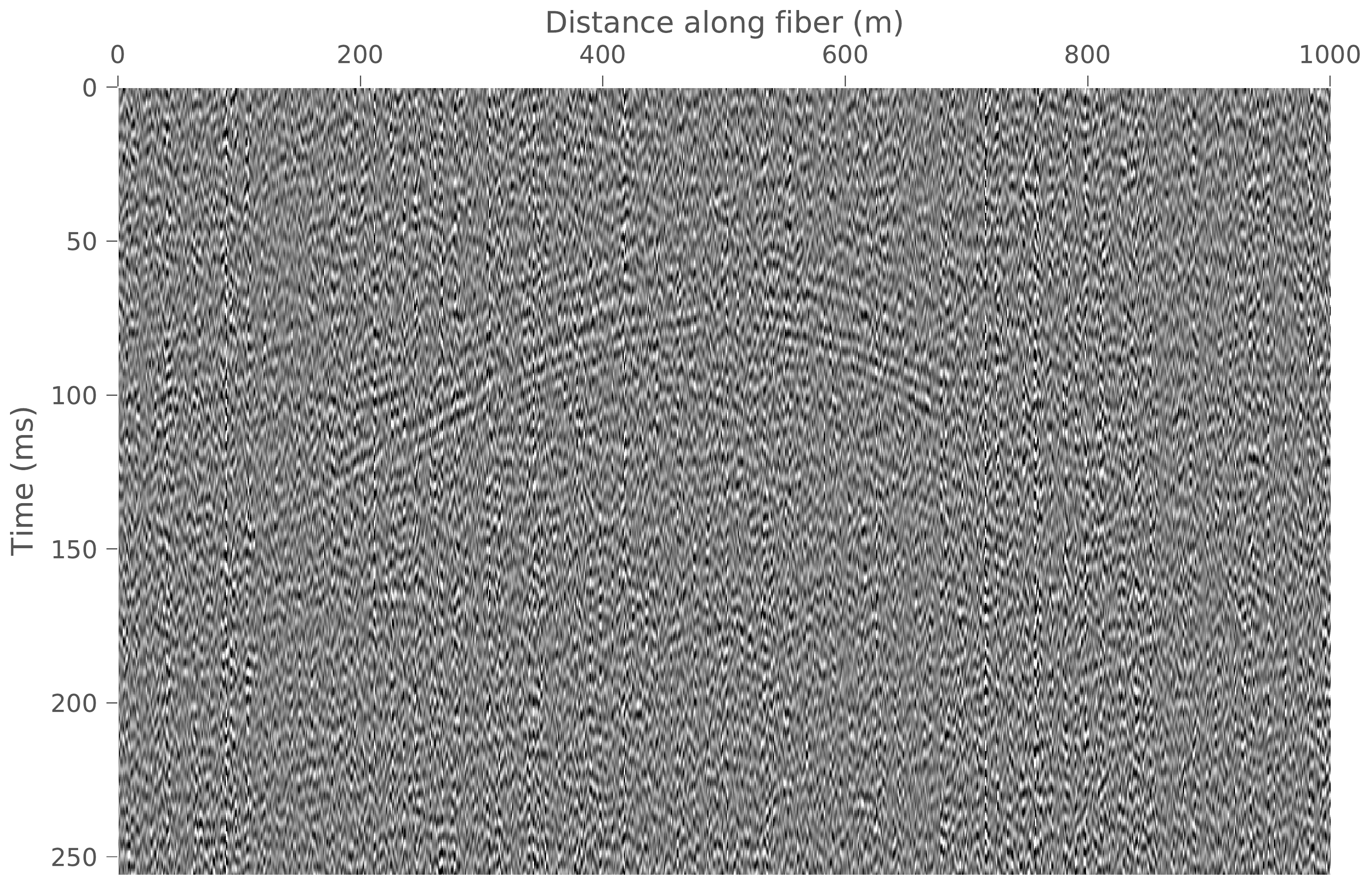}
    \includegraphics[width=0.42\linewidth]{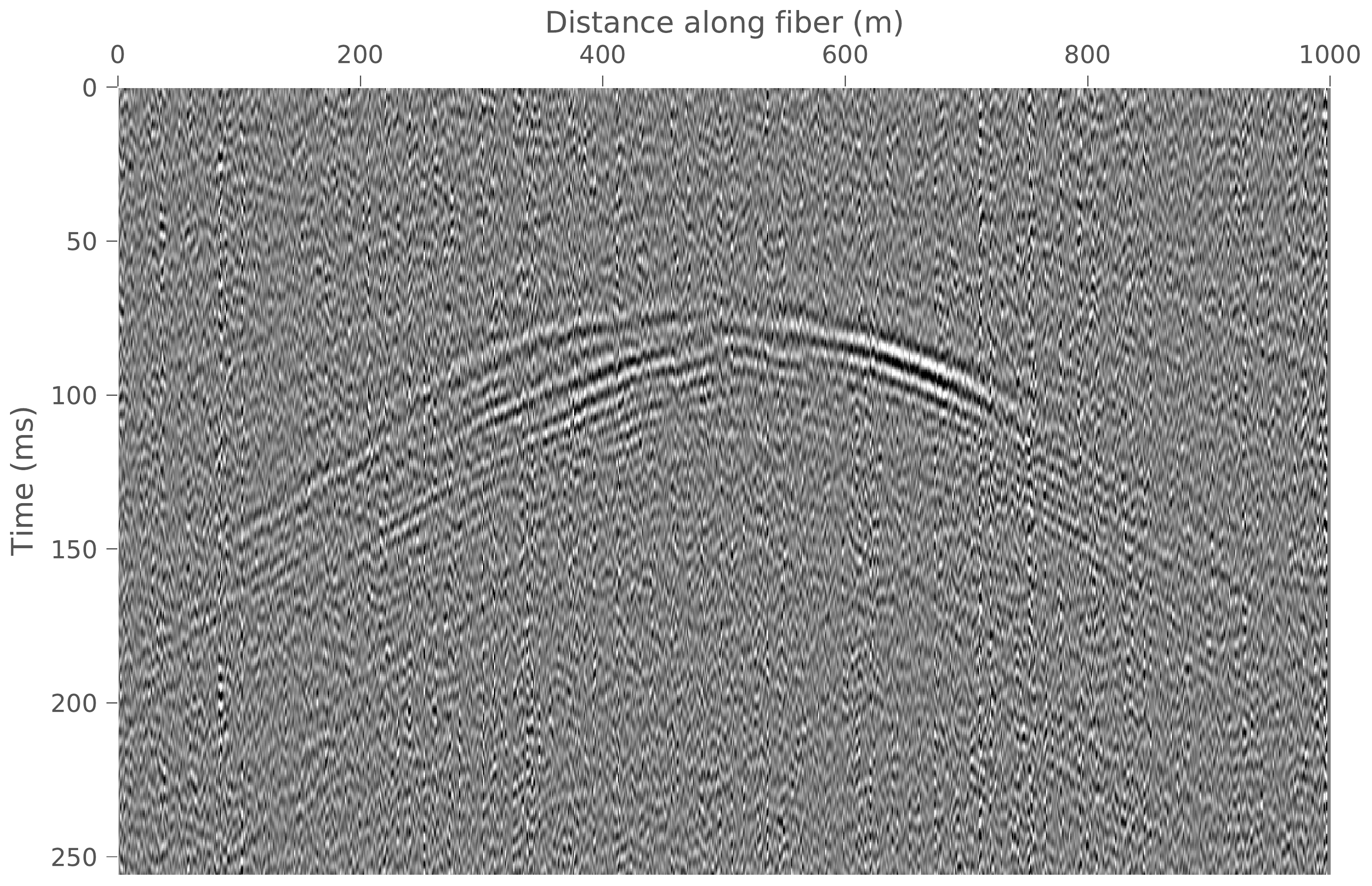}
    \includegraphics[width=0.42\linewidth]{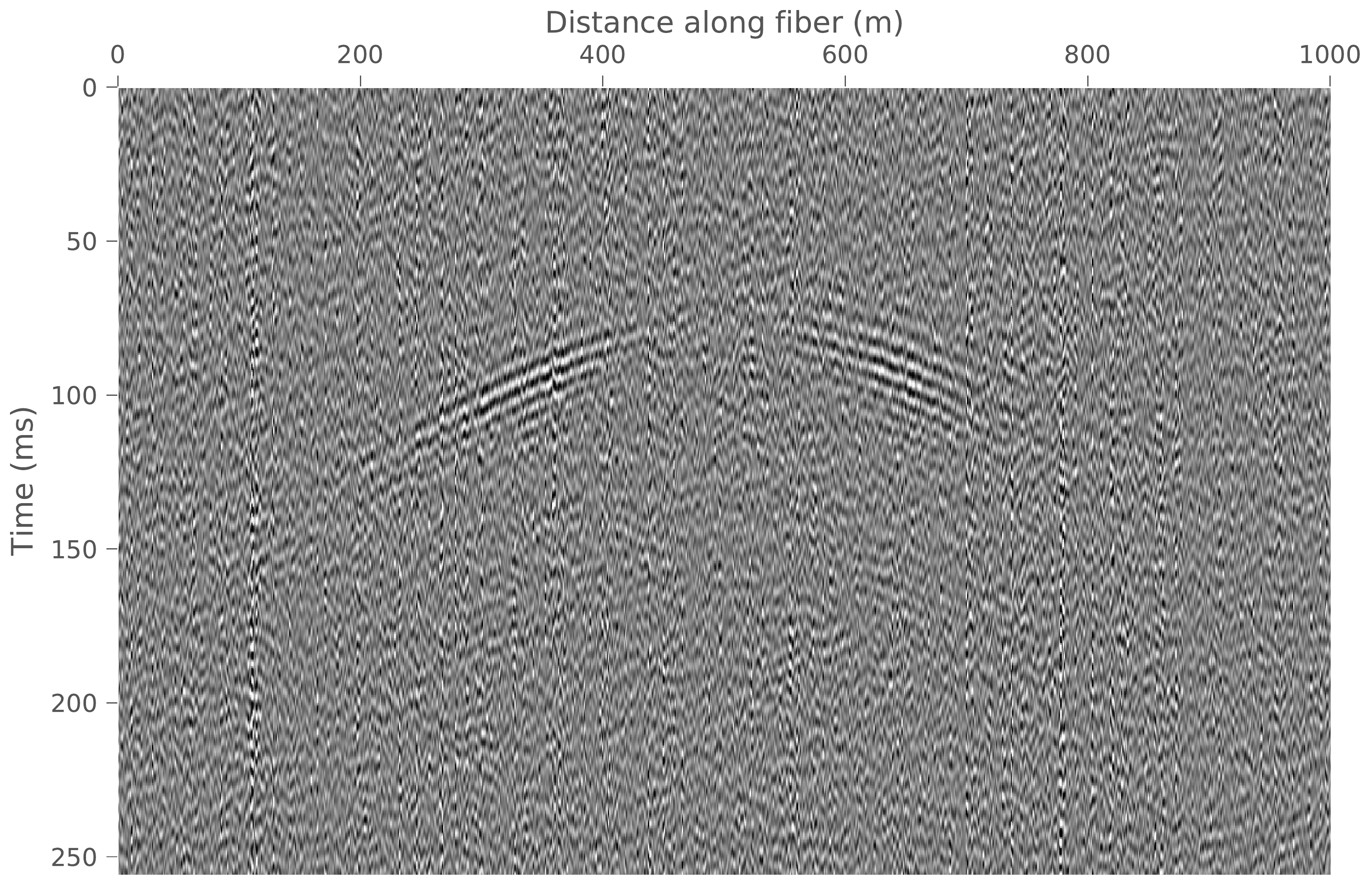}
    \end{center}
    \caption{Microseismic events recorded over the DAS fiber. While some events appear clearly in the data, many are faint and have similar amplitude as the background noise, making them unsuitable for standard channel-by-channel thresholding detection methods. 
    \label{example_events}}
\end{figure}

To train a deep learning classifier to detect microseismic events, we create a curated catalog of  events and background noise examples. We manually picked all the microseismic events over two stimulation stages, for each of which we analyzed an hour of data, resulting in almost 7,000 events (Figures \ref{example_events} and \ref{stage_12}). The manual picking was performed by two geophysicists: a graduate student and a post-doctoral researcher. The events were picked by visual inspection on continuous data, band-pass filtered between 10 and 200$\,$Hz, a frequency range that contains most of the microseismic energy. Eventually, the entire catalog was inspected for errors by three geophysicists.

We select random time windows at least 30 seconds away from any microseismic event over the two hours of manually-labeled data for examples of background noise. Since the events tend to occur in swarms during the hydraulic stimulation stage, this approach ensures these background noise windows are unlikely to contain any low-amplitude event we might have missed during picking. The noise windows were then visually inspected to ensure that they did not contain any microseismic energy.

\subsection{Data processing}

To remove the common-mode noise, we first subtract the median over all the channels at each time sample. We then band-pass filter the data between 10 and 200$\,$Hz, and to reduce computational memory load, we decimate the data to 500 samples per second and store them as 32-bit floating-point numbers. Since seismic data have an extensive dynamic range, we clip the data at the 99.5th percentile of the absolute amplitudes to minimize the impact of extreme values. The data are already centered around zero after the filter. We then divide each channel by its standard deviation, computed after clipping. We calculate these data statistics (percentiles and standard deviations) over ten stimulation stages, representing about 20 hours of continuous data.

\begin{figure}
\begin{center}
\includegraphics[width=0.025\linewidth,trim={9cm 11.5cm 8.5cm 9.5cm},clip]{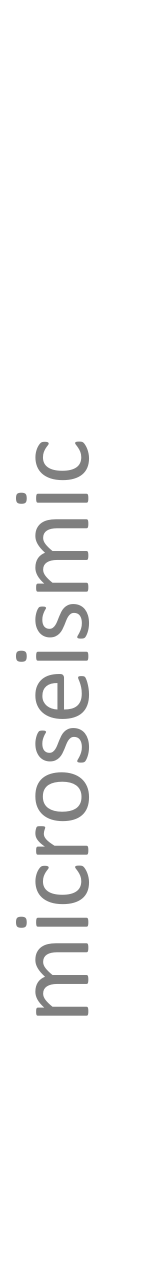}
\includegraphics[width=0.31\linewidth]{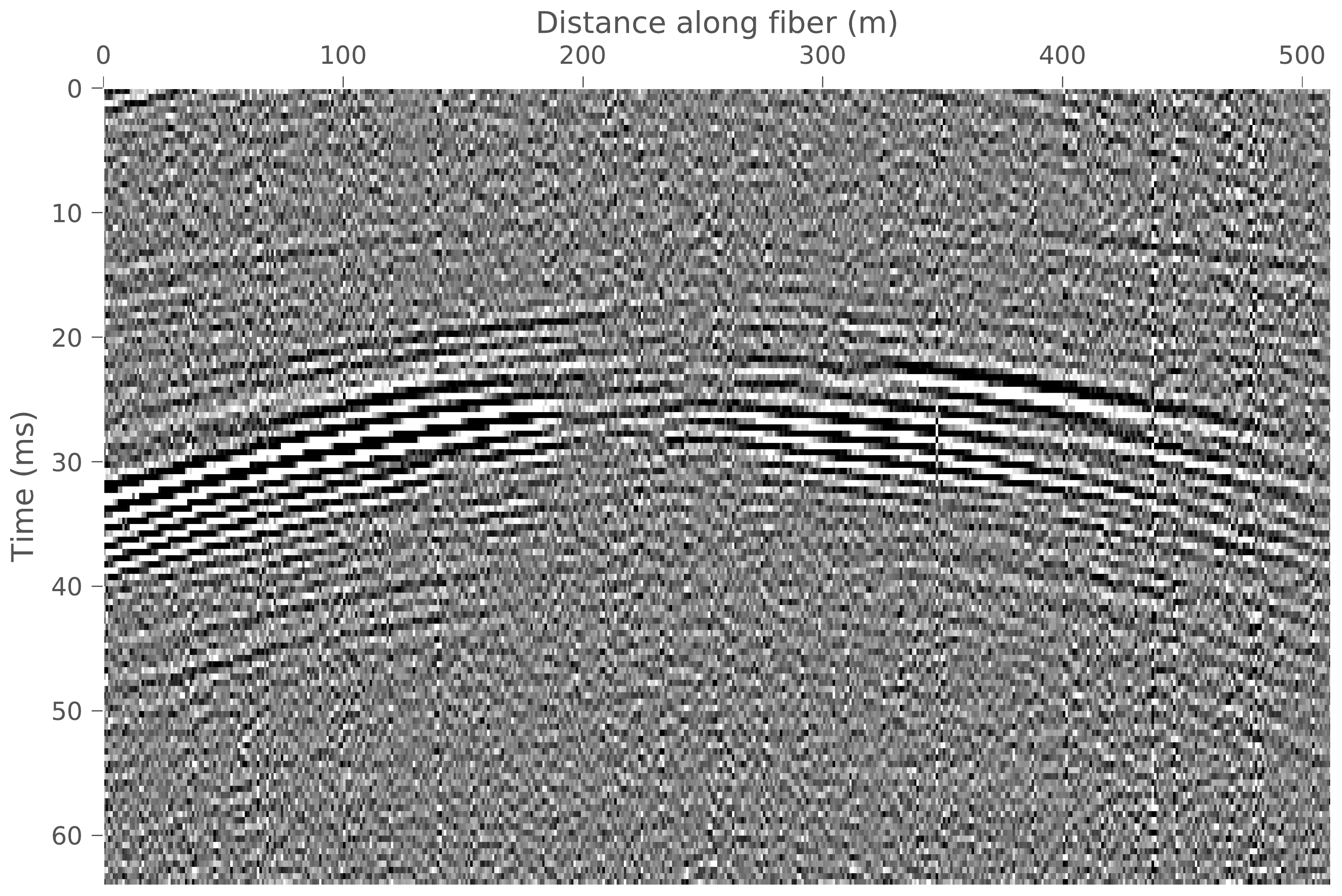}
\includegraphics[width=0.31\linewidth]{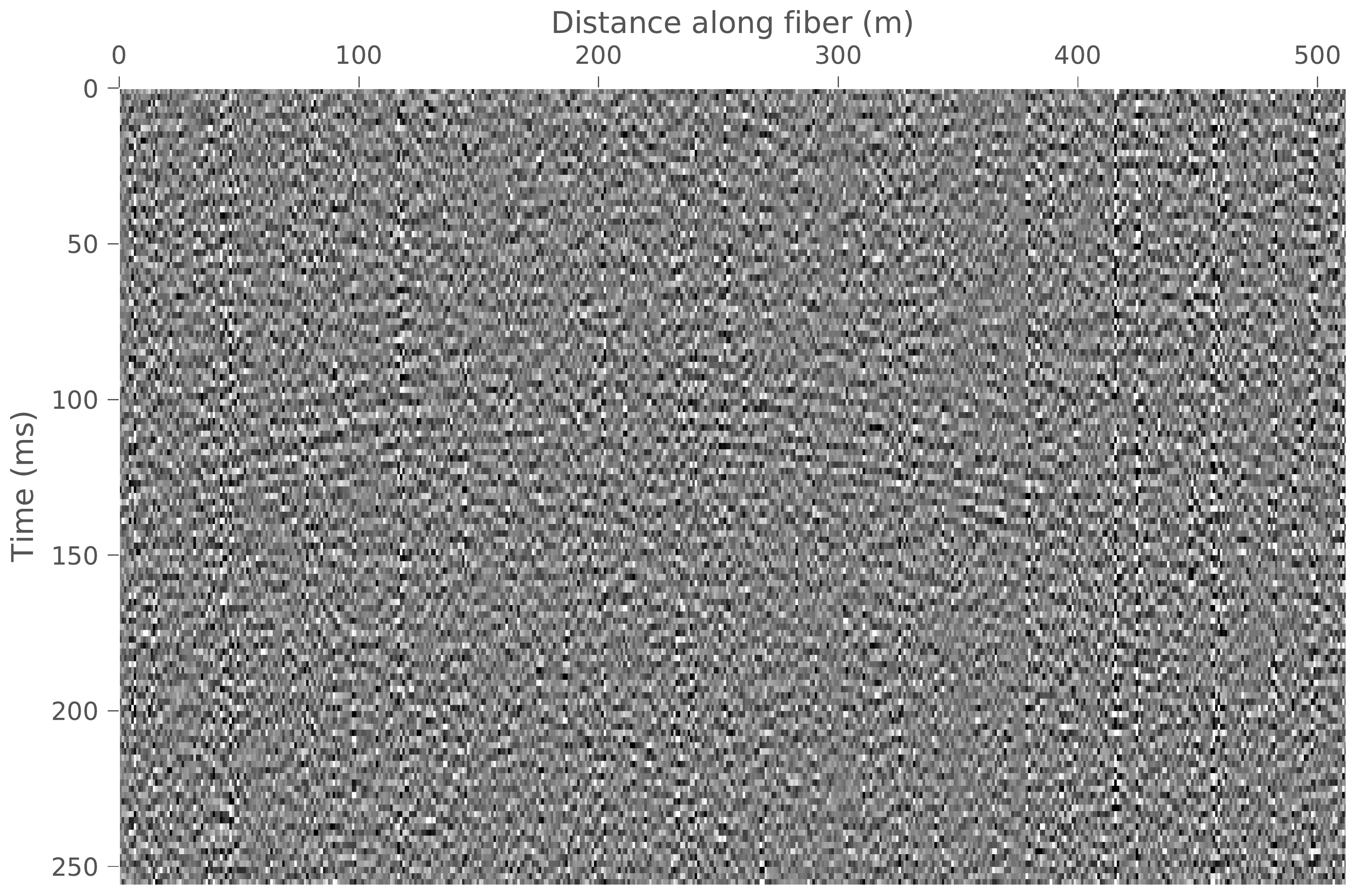}
\includegraphics[width=0.31\linewidth]{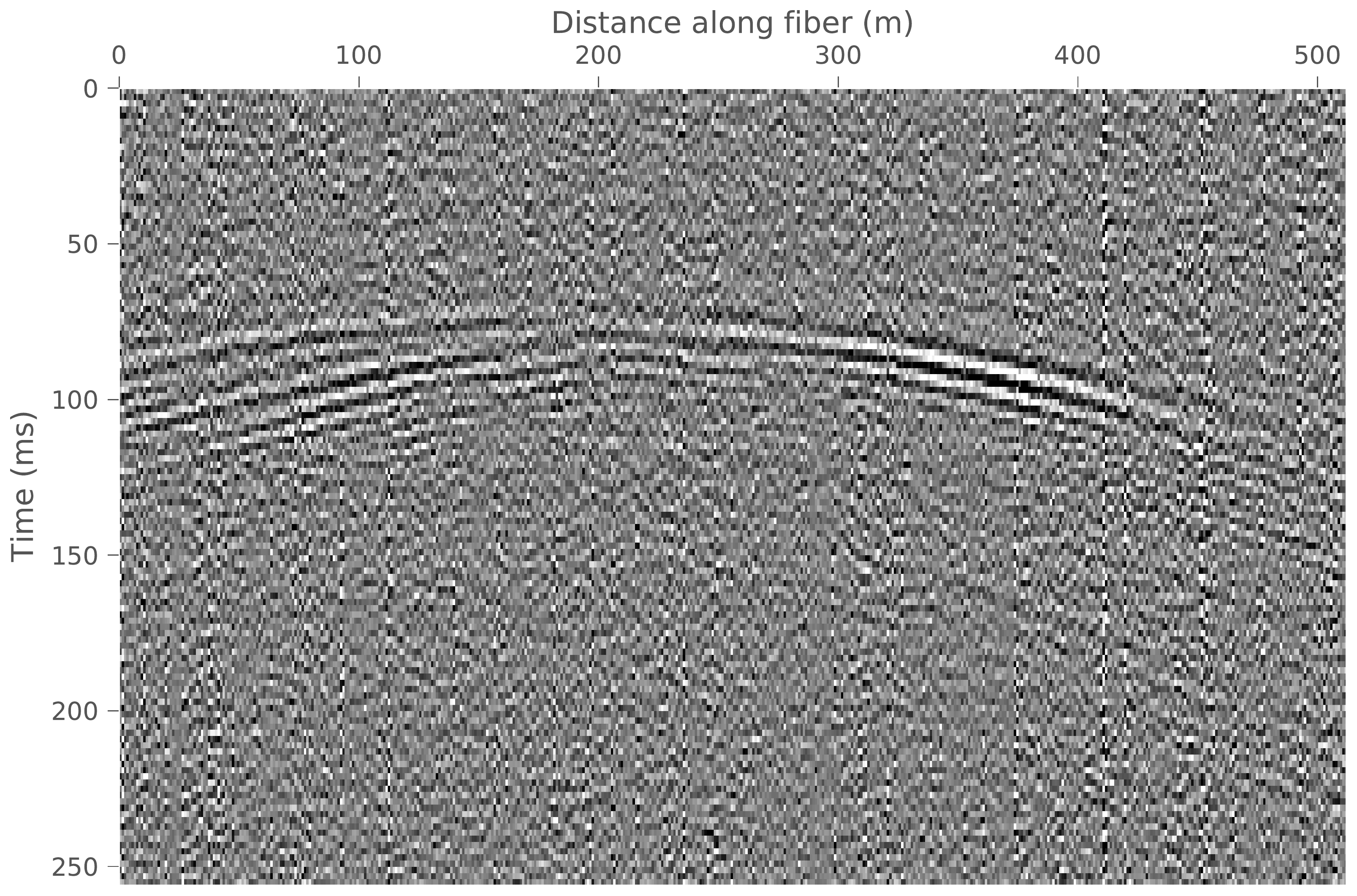}

\vspace{0.3cm}

\includegraphics[width=0.025\linewidth,trim={9cm 11.5cm 8.5cm 9.5cm},clip]{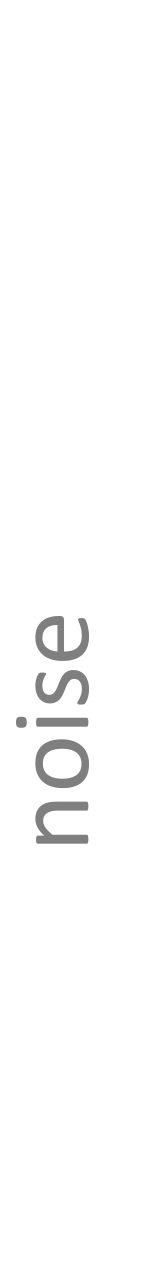}
\includegraphics[width=0.31\linewidth]{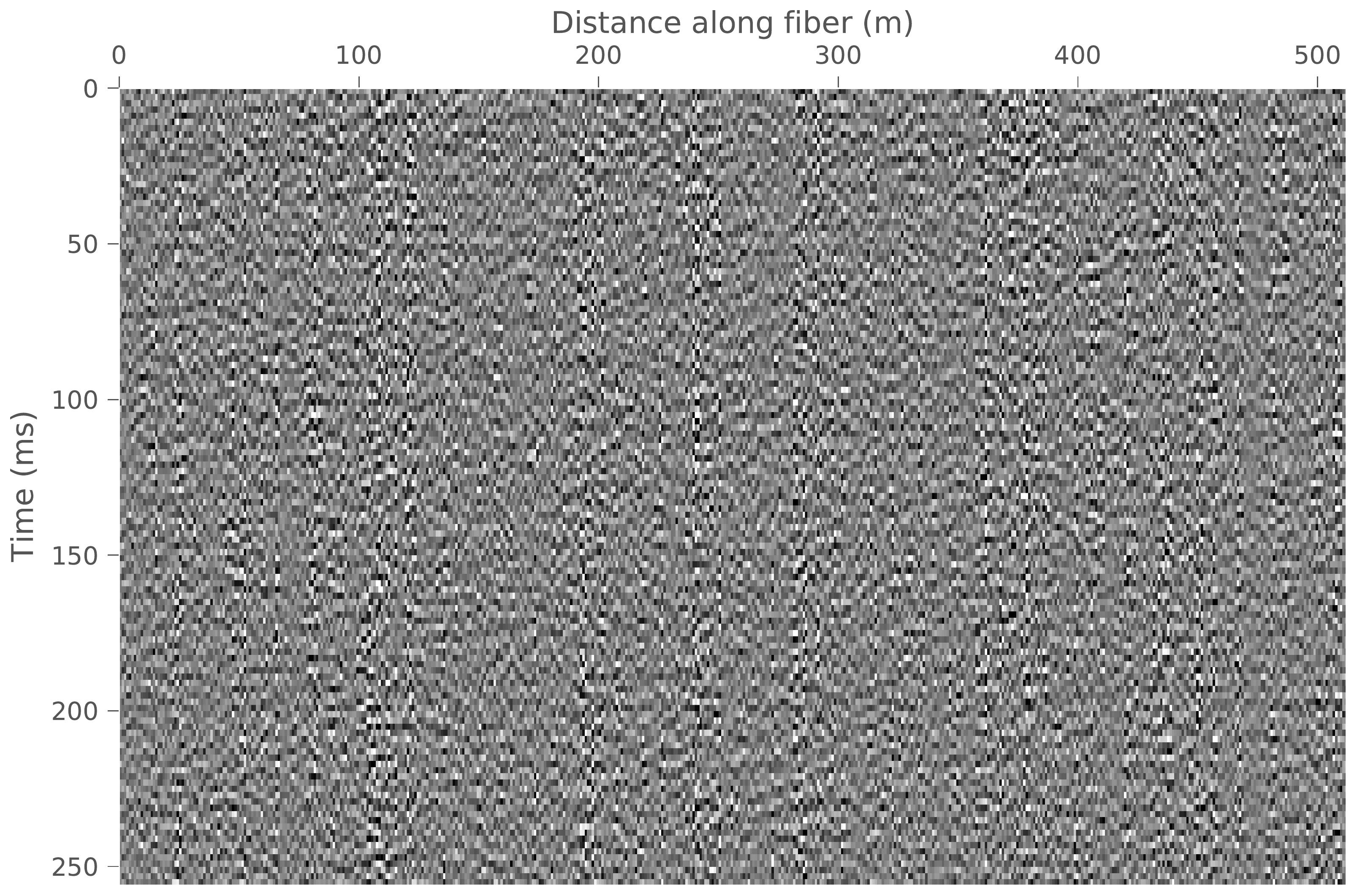}
\includegraphics[width=0.31\linewidth]{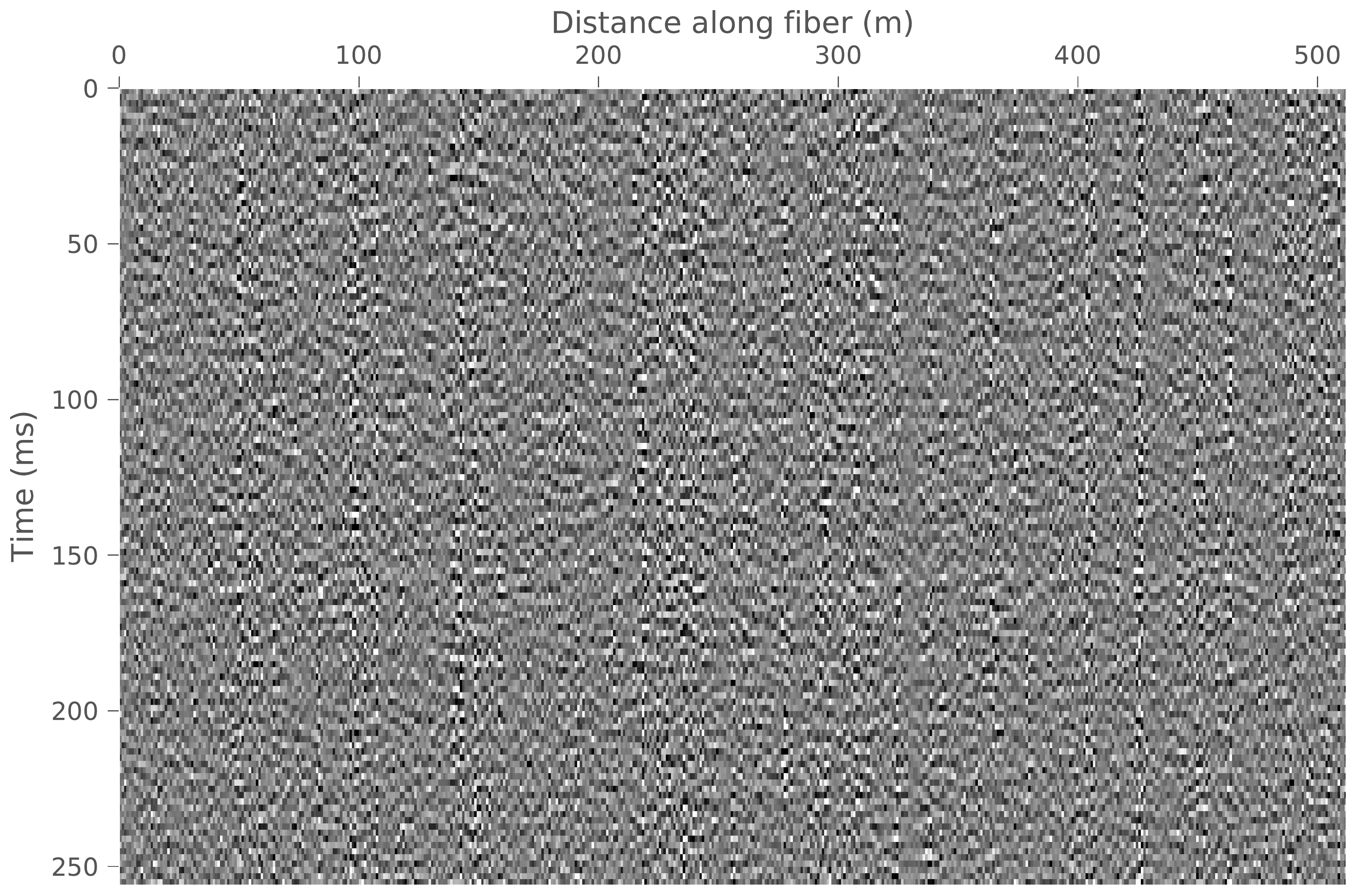}
\includegraphics[width=0.31\linewidth]{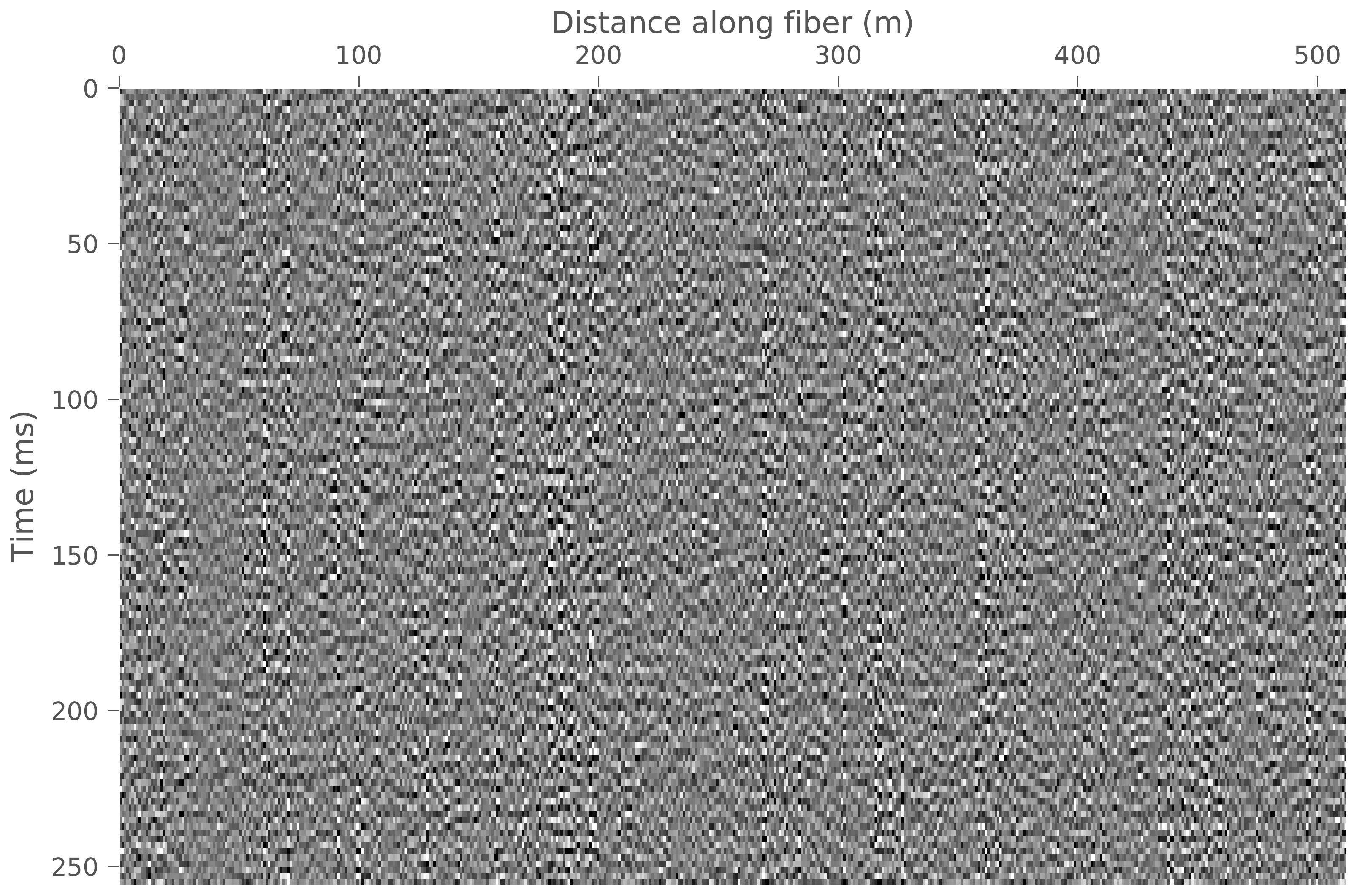}
\end{center}
\caption{Processed data windows used as inputs for the machine learning classifier. The top row presents examples of microseismic events of varying amplitudes; the bottom row shows background noise.
\label{input_data}}
\end{figure}

We prepare two-dimensional (channel $\times$ time) data windows, labeled as events if they contain a microseismic event or as noise otherwise. We extract the microseismic data windows such as to include the apex of the events, since we expect to have full wavefields for all the stages along the fiber. We use a detection window of 512 channels and 0.256 seconds (128 time samples). For data augmentation, we feed the data as larger windows (700 channels $\times$ 196 time samples), from which we extract random windows of the size of the detection window (512 channels $\times$ 128 time samples) in the machine learning input pipeline. Examples of the processed data windows used as input for the machine learning classifier are presented in Figure~\ref{input_data}. 

We split the data into training, evaluation, and test sets with a ratio of 8:1:1  while maintaining a 1:1 balance between events and background noise samples (Table \ref{dataset}). We sort all the data samples by time stamp before splitting to ensure no overlapping windows between the datasets.

\begin{table}
    \renewcommand{\arraystretch}{1.4}
    \center
    \caption{Number of microseismic and noise data windows in the training, evaluation, and test set.}
    \label{dataset} 
    \begin{tabular*}{\columnwidth}{c @{\extracolsep{\fill}} cccc}
                    & \bf Training  &  \bf Evaluation    &  \bf Test  \\
        \hline
        \hline
Microseismic   & 5,430   & 678 &  678   \\
Noise      & 5,430   & 678 &  678    \\
        \hline
        Total       & 10,860  & 1,356 & 1,356   \\
    \end{tabular*}
\end{table}

\section{Machine learning model}
\label{sec:hptuning}

We frame the machine learning task as a binary classification between data windows containing a microseismic event and those containing background noise. We design our workflow to optimize the network architecture and the training hyperparameters jointly.

We create two modular networks, a VGG-type one and a ResNet-type one, and use a small number of hyperparameters to control the architecture details. VGG and ResNet are the two neural networks that achieved top performance at the ImageNet Challenge 2014 and 2015, respectively (\cite{simonyan2015deep, he2015deep,ILSVRC15}). 
This modular approach allows us to experiment with networks that are similar to these models but with a smaller number of layers (\cite{huot2021detecting}). Indeed, our microseismic detection task only involves  two classification categories, namely microseismic and noise, compared to the thousand categories of the ImageNet Challenge. Therefore, the number of data representations that the neural network has to capture can be much smaller, suggesting that a smaller network might be sufficient to tackle these tasks. 

Deep networks are notoriously hard to train because of the vanishing gradient problem — as the gradient is back-propagated to earlier layers, repeated multiplication may make it infinitely small. Additionally, the high non-linearity of the network increases the chances of getting stuck at local minima. Smaller neural networks have the advantage of being easier to train and require less computational resources, making them more cost-effective. They are faster to train and evaluate, reducing the overall time of the machine learning cycle, allowing for faster operational decisions. Most importantly, smaller networks are less prone to overfitting (\cite{goodfellow2016deep}), which is particularly relevant in geophysics, where labeled datasets are often curated by domain experts and therefore limited in size. Their lower representation power also makes them more robust to noise in the data.

\subsection{Modular architectures}
\label{sec:modular_architectures}

We define the VGG-type network as a modular sequence of convolutional blocks and the ResNet-type network as a modular sequence of residual blocks. 
The residual blocks have shortcut connections, where the input of the block is added to the output of the block. 
All the convolution layers have $3 \times 3$ kernels and $1 \times 1$ strides as proposed by \cite{simonyan2015deep}. At the output of the VGG-type network, we add two fully connected layers with activation, and at the output of the ResNet-type network, we use a layer of global average pooling with activation, as described by \cite{simonyan2015deep} and \cite{he2015deep}. 
Both networks then have a fully-connected output layer from which we compute the loss. We fine-tune the network architecture using the following hyperparameters:

\begin{itemize}
    \item \textbf{Number of convolutional or residual blocks,} selected from 2 to 5. Please note that with this definition, our ResNet-type is deeper than our VGG-type network. 
    \item \textbf{Number of filters in each block,} constrained to be powers of 2. We control the number of filters with three hyperparameters: the number of filters in the first block, how we increase the number of filters in the following blocks, and the number of filters in the last layer of the network. The number of filters in the first block is set from 16 to 64. The number of filters can then increase either linearly, double every other block, or double every block.  When the number of filters increases in the next block, we add a downsampling layer between the two blocks.
    \item \textbf{Type of activation,} selected between ReLU or Leaky ReLU (\cite{ramachandran2017searching}). Both ReLU and Leaky ReLU are widely-used activation functions (\cite{goodfellow2016deep}). ReLU tends to promote sparsity by removing all negative values, while Leaky ReLU still allows negative values.
    \item \textbf{Type of downsampling,} either by $2 \times 2$ Max Pooling or by strided convolution with a $3 \times 3$ kernel and $2 \times 2$ stride. Max Pooling promotes sparsity, while the strided convolution is a softer form of downsampling.
    \item \textbf{Whether to use batch normalization in the blocks} (\cite{ioffe2015batch}).
    \item \textbf{The dropout rate to use in the convolutional blocks,} selected between 0, 10\%, or 20\%. 
    This hyperparameter is ignored for the residual blocks.
    \item \textbf{The dropout rate to use at the end of the network,} selected between 0 and 80\%, in 10\% increments.
    \item \textbf{How to perform regularization}:
    \begin{itemize}
        \item \textbf{Type of regularization on the weights,} constrained to none, $L_1$ or $L_2$.
        \item \textbf{Regularization weight,}  selected from $10^{-5}$ to $10^{-3}$.
    \end{itemize}
\end{itemize}

In addition to the hyperparameters that control the architecture, we define the following hyperparameters related to the training:
\begin{itemize}
\item \textbf{Learning rate,} selected from $10^{-5}$ to $10^{-2}$.
\item \textbf{Batch size,} selected between 32 and 128 in powers of 2.
\end{itemize}

The neural networks are implemented in TensorFlow (\cite{tensorflow2015-whitepaper}). We use the cross-entropy loss as loss function. While there is no guarantee that these design choices result in an optimal network, the modular implementation allows us to easily experiment with a  range of architectures. These networks are designed to re-use many of the learnings of state-of-the-art models, making these networks reasonable directions to explore.

\subsection{Hyperparameter tuning}
A common approach to hyperparameter selection is grid search, which performs an exhaustive search of the hyperparameter space. It is simple to implement and, being embarrassingly parallel, straightforward to parallelize over many computational ressources. However, it suffers from the curse of dimensionality, making it untractable for larger numbers of hyperparameters. To overcome this limitation, a conventional algorithm for hyperparameter selection is random search, which evaluates the performance of random combinations of hyperparameters. While random search may perform well on specific problems, it does not use the information provided by previous experiments to select the next set of hyperparameters to evaluate. This behavior is generally undesirable when the cost of training and evaluating a neural network is high. In our machine learning workflow, we implement the hyperparameter search using Bayesian optimization (\cite{snoek2012practical}), a method for the global optimization of functions with expensive evaluations. We use the GPyOpt library (\citet{gpyopt2016}) for the implementation of the Gaussian Process and use the loss computed on the evaluation dataset as our performance metric.
To improve the convergence of the Bayesian optimization, we rescale the hyperparameter domain by bring all the values to a similar range:
\begin{itemize}
\item For hyperparameters such as the number of filters and the batch size, we take the base 2 logarithm.
\item For hyperparameters such as the learning rate and the regularization weight, we take the base 10 logarithm. 
\item For hyperparameters such as the dropout rates, we rescale the decimals to integer values.
\item All other hyperparameters are treated as categorical one-hot encodings. 
\end{itemize}

\begin{figure}
    \begin{center}
    \includegraphics[width=\linewidth,trim={3cm 7cm 1cm 6.5cm},clip]{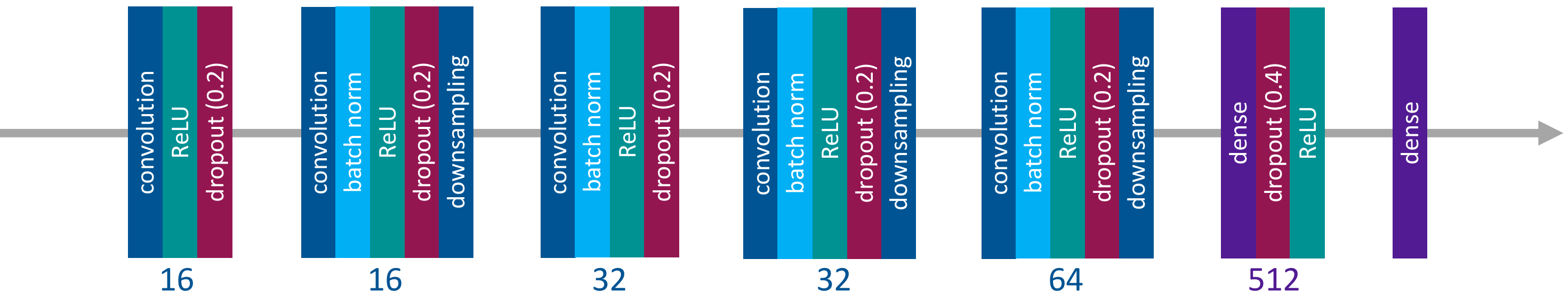}
    \end{center}
    \caption{Our best-performing network architecture for microseismic event detection obtained by hyperparameter tuning. The convolution layers have a 3 $\times$ 3 kernel and a 1  $\times$ 1 stride. The downsampling is performed by strided convolutions with a 3 $\times$ 3 kernel and 2  $\times$ 2  stride. The number of filters is indicated under each block. The dropout rate is in parenthesis for each dropout layer. 
    } 
    \label{network}
\end{figure}

For each iteration of Bayesian optimization, we train the network with Adam optimizer (\cite{kingma2014adam}) for 100 epochs with early-stopping if the network performance does not improve over ten epochs, using two NVIDIA V100 GPUs on an on-premise cluster. As shown in Figure \ref{network}, our best-performing model is a VGG-type model with five convolutional blocks.
It uses a learning rate of 0.0001 and a batch size of 32.

\begin{figure}
\begin{center}
\includegraphics[width=0.6\linewidth]{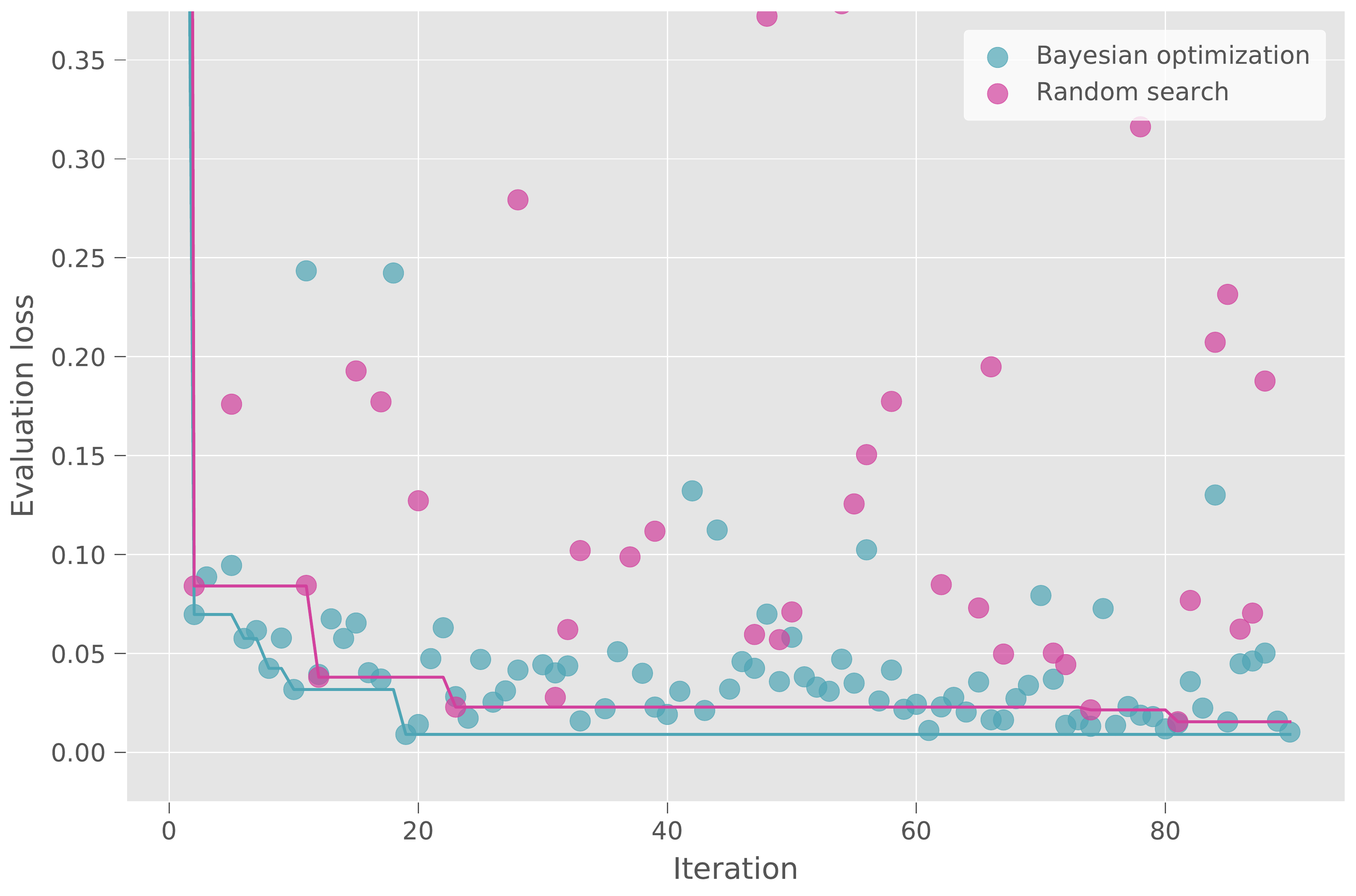}
\end{center}
\caption{Hyperparameter optimization convergence over multiple experiments for microseismic event detection. For each iteration, we train the neural network for a candidate set of hyperparameters and use the loss computed over the evaluation dataset as our performance metric. Every dot represents an experiment, while the solid line shows the convergence of the algorithm by keeping track of the best-performing model obtained thus far. 
\label{hptuning_convergence}}
\end{figure}

To validate the convergence of our hyperparameter tuning, we compare it to a random search of the hyperparameter space (Figure \ref{hptuning_convergence}). We see that our Bayesian optimization approach samples promising hyperparameters with a much higher frequency than random search and yields a better-performing network over 90 experiments.

We note that our five best-performing networks are all VGG-type networks that share many similarities: same number of convolutional blocks, same learning rate, downsampling by strided convolution, no regularization. 
These are relatively small networks compared to state-of-the-art classifiers. In our experiment, deeper networks tended to overfit our training data. We could address this problem by expanding the dataset, but given that microseismic events have to be manually picked and verified, this would come at the cost of time and resources. 

\section{Experimental results}

\subsection{Benchmark analysis}
\label{sec:benchmark}

Our trained machine learning model achieves 98.6\% accuracy on our benchmark test dataset, with a precision of 99.92\% and a recall of 97.58\% (Table \ref{metrics}). This result shows that our model performs remarkably well at distinguishing microseismic events from background noise despite the smaller network. It suggests that small neural networks can automate specialized tasks with high accuracy, which is particularly relevant in geophysics, where labeled datasets are often curated by domain experts and therefore limited in size. 

\begin{table}
    \renewcommand{\arraystretch}{1.4}
    \begin{center}
    \caption{Classification metrics for microseismic event detection computed against our benchmark test dataset. }
    \label{metrics}
    \begin{tabular*}{\columnwidth}{c @{\extracolsep{\fill}} ccccc}
                    & \bf Accuracy & \bf AUC\textsuperscript{*} & \bf Precision & \bf Recall \\
        \hline 
        \hline
        Benchmark test\textsuperscript{$\dagger$}          & 98.68\% & 99.72\% & 99.92\% & 97.58\% \\
        Mean (Standard error)\textsuperscript{$\dagger\dagger$}   & 98.59\% (0.16) & 99.73\% (0.03) & 99.33\% (0.15)& 97.86\% (0.27)\\
    \end{tabular*}
    \end{center}
    \footnotesize{\textsuperscript{*}Area under the curve. \\
    \textsuperscript{$\dagger$}Metrics obtained with the best-performing model. \\
    \textsuperscript{$\dagger\dagger$}Mean and standard error for each of the metrics obtained by re-training and re-evaluating the model five times.}
\end{table}

\begin{figure}
    \begin{center}
    \includegraphics[width=0.48\linewidth]{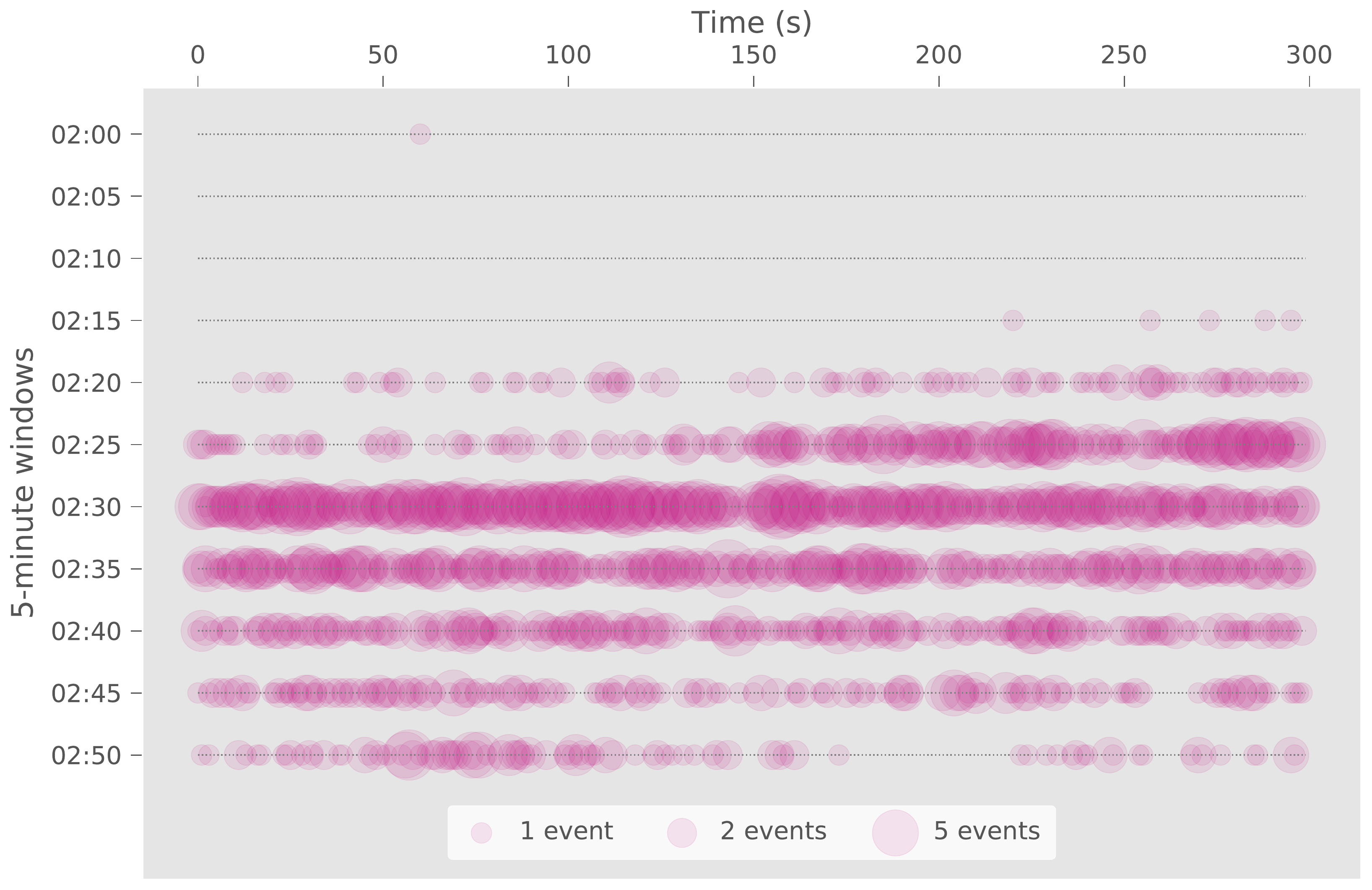}
    \includegraphics[width=0.48\linewidth]{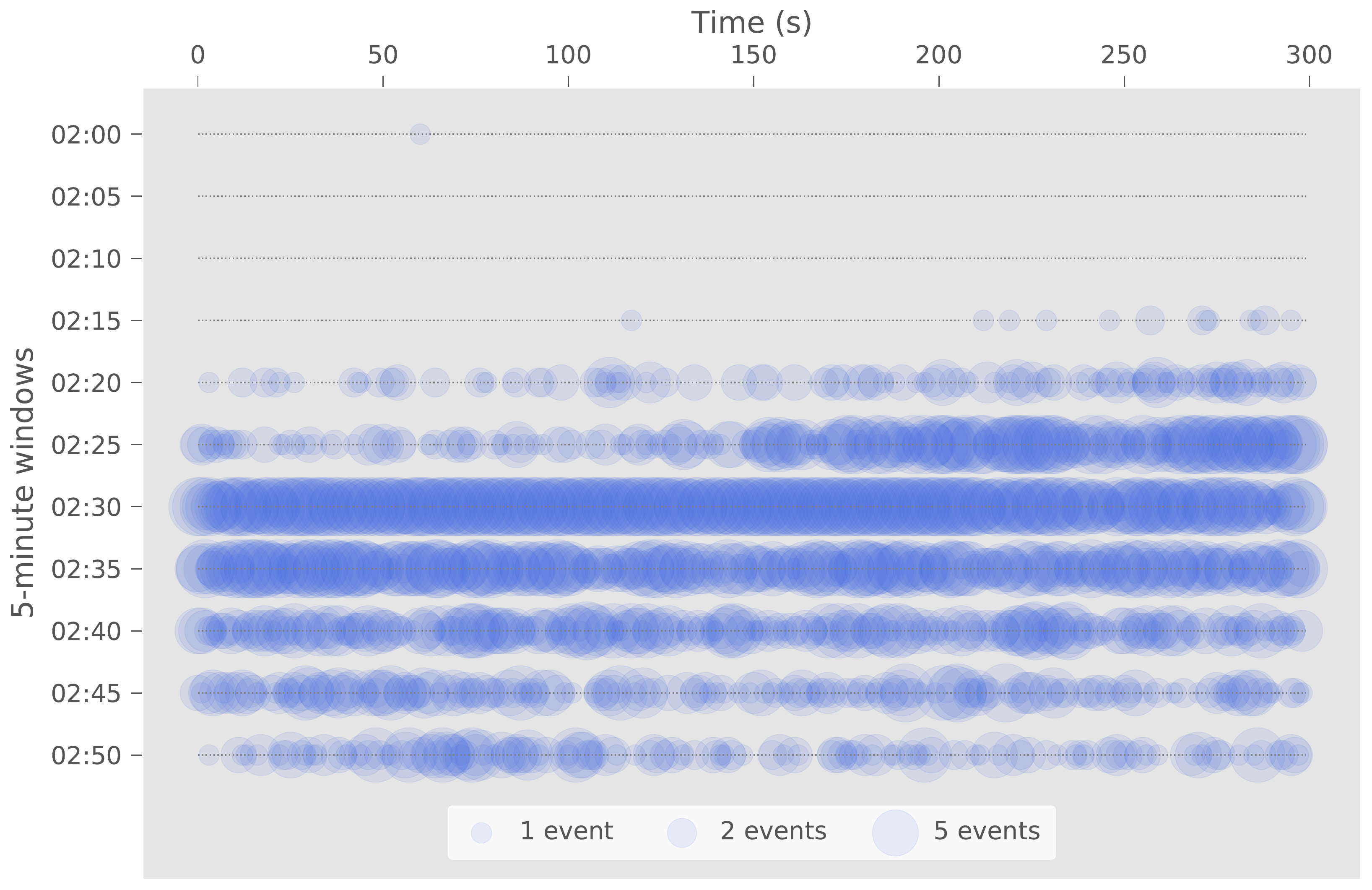}
    \end{center}
    \caption{Microseismic event density for a stimulation stage on which we manually picked all the microseismic events. Each line represents 5 minutes of data, and every circle corresponds to the number of events per second. (Left) Manually-picked event catalog. The number of events per second varies between 0 and 10. (Right) Machine learning model predictions for the same stage. 
    }
    \label{stage_12}
\end{figure}

We validate our model's performance over continuous data by running it over one of the manually-labeled stages. We scale the model to continuous data using a sliding detection window with a stride of 125 ms and classifying each detection window independently. Figure \ref{stage_12} shows that the network yields similar detections compared to the catalog. The detection window size limits the detection capabilities since the model does not distinguish between events when there are multiple within a single window of 256 ms. However, this limitation is not a problem for fracture characterization and monitoring, for which we do not require the exact number of events. 

\begin{figure}
    \begin{center}
    \includegraphics[width=0.49\linewidth]{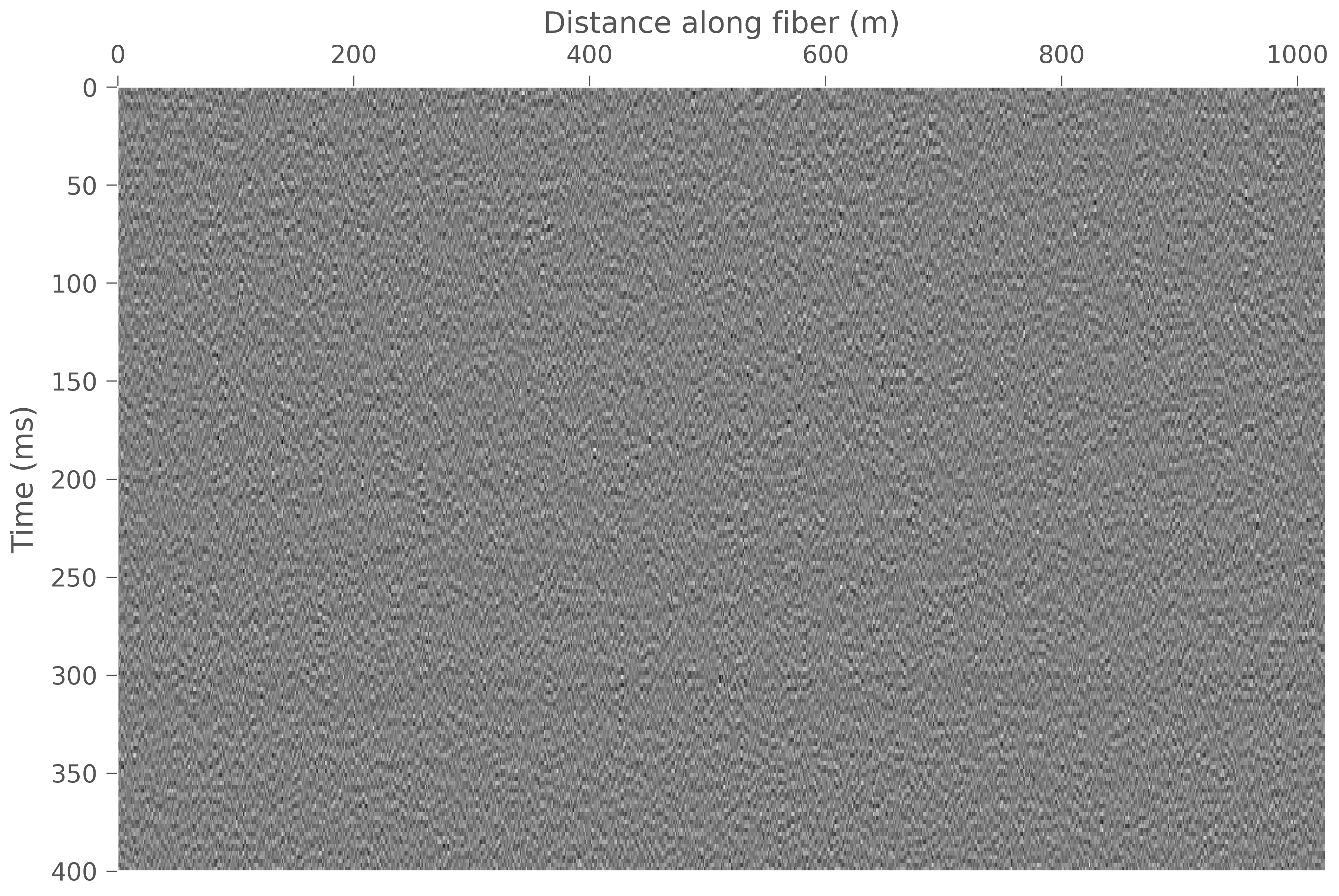}
    \includegraphics[width=0.49\linewidth]{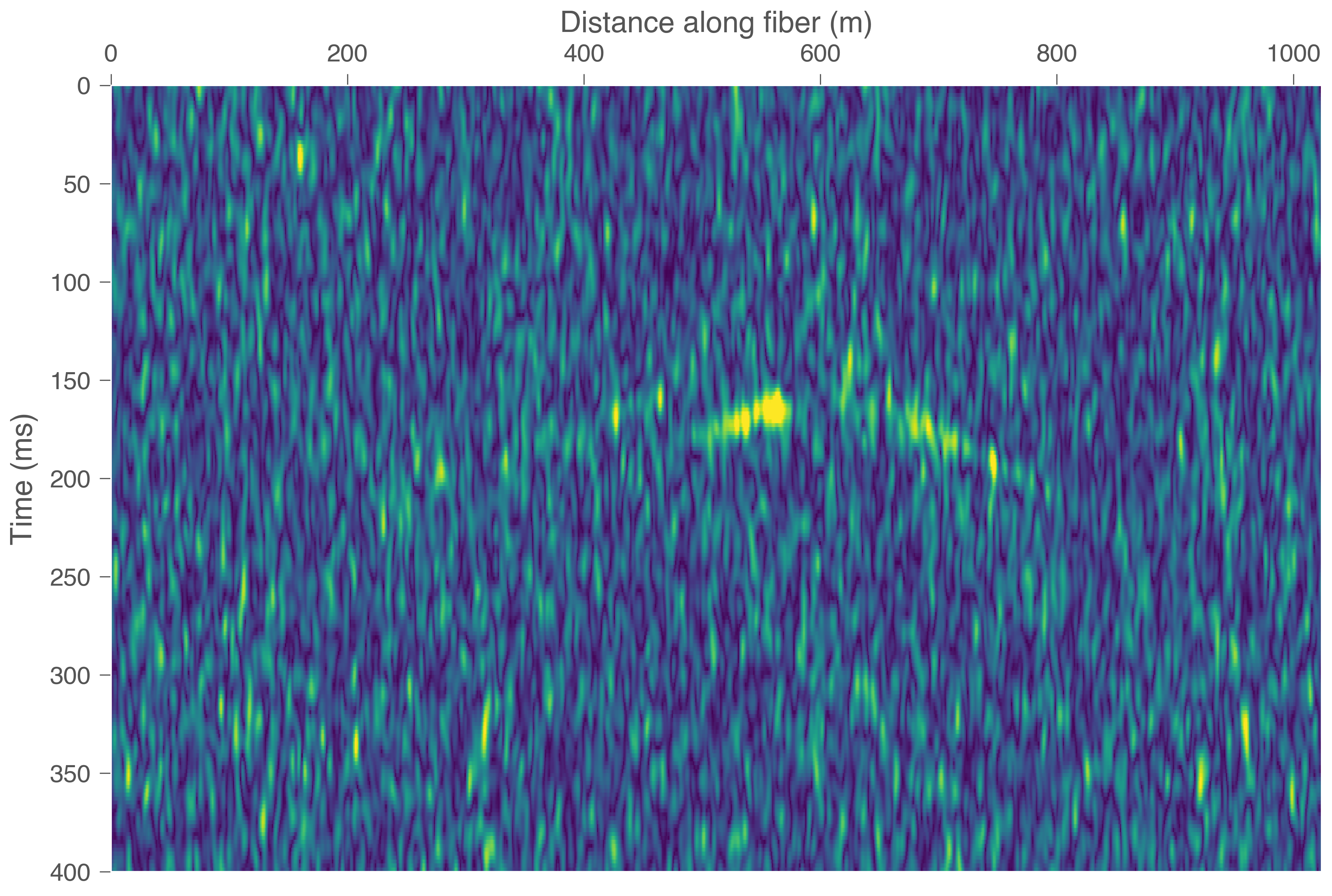}
    \includegraphics[width=0.49\linewidth]{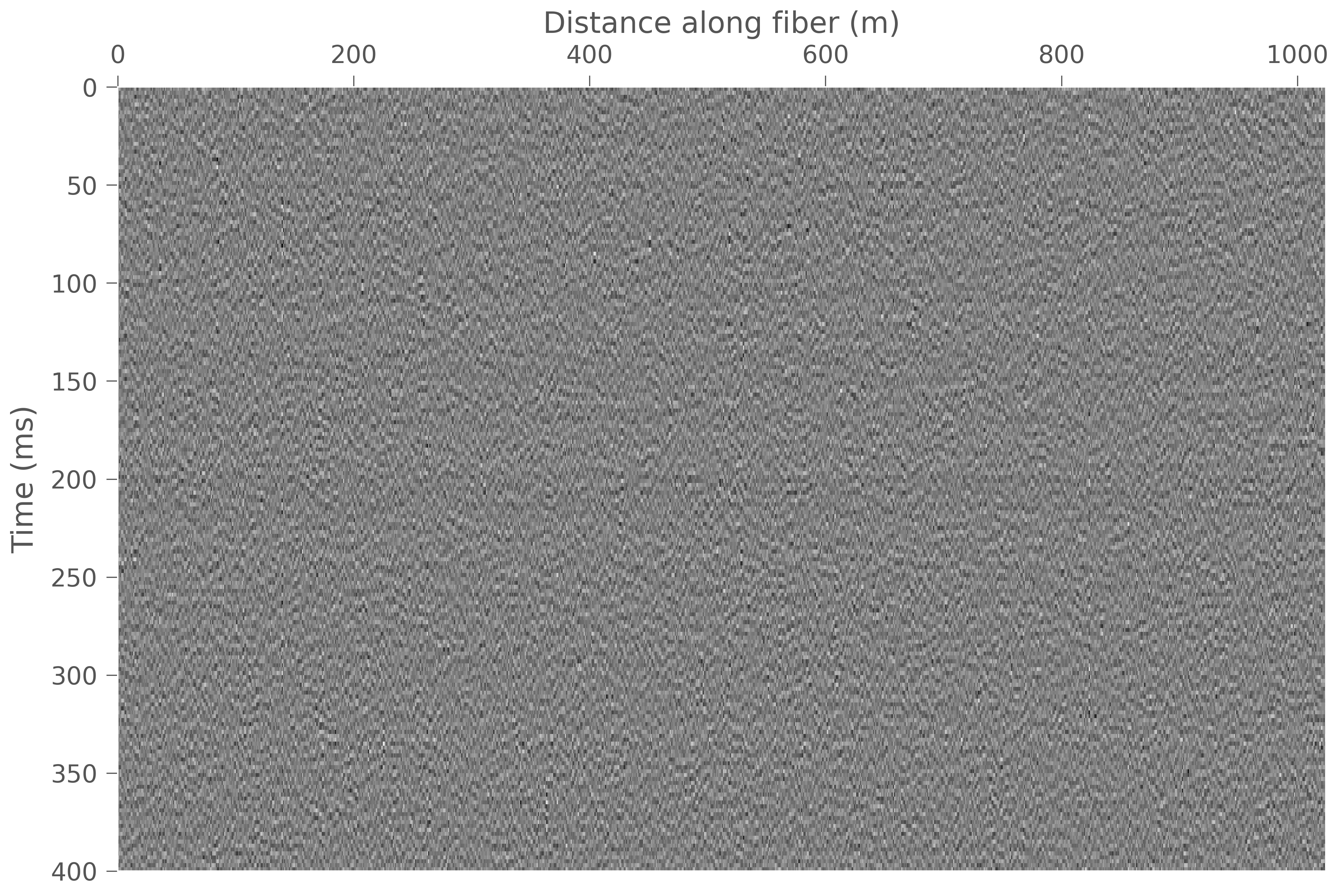}
    \includegraphics[width=0.49\linewidth]{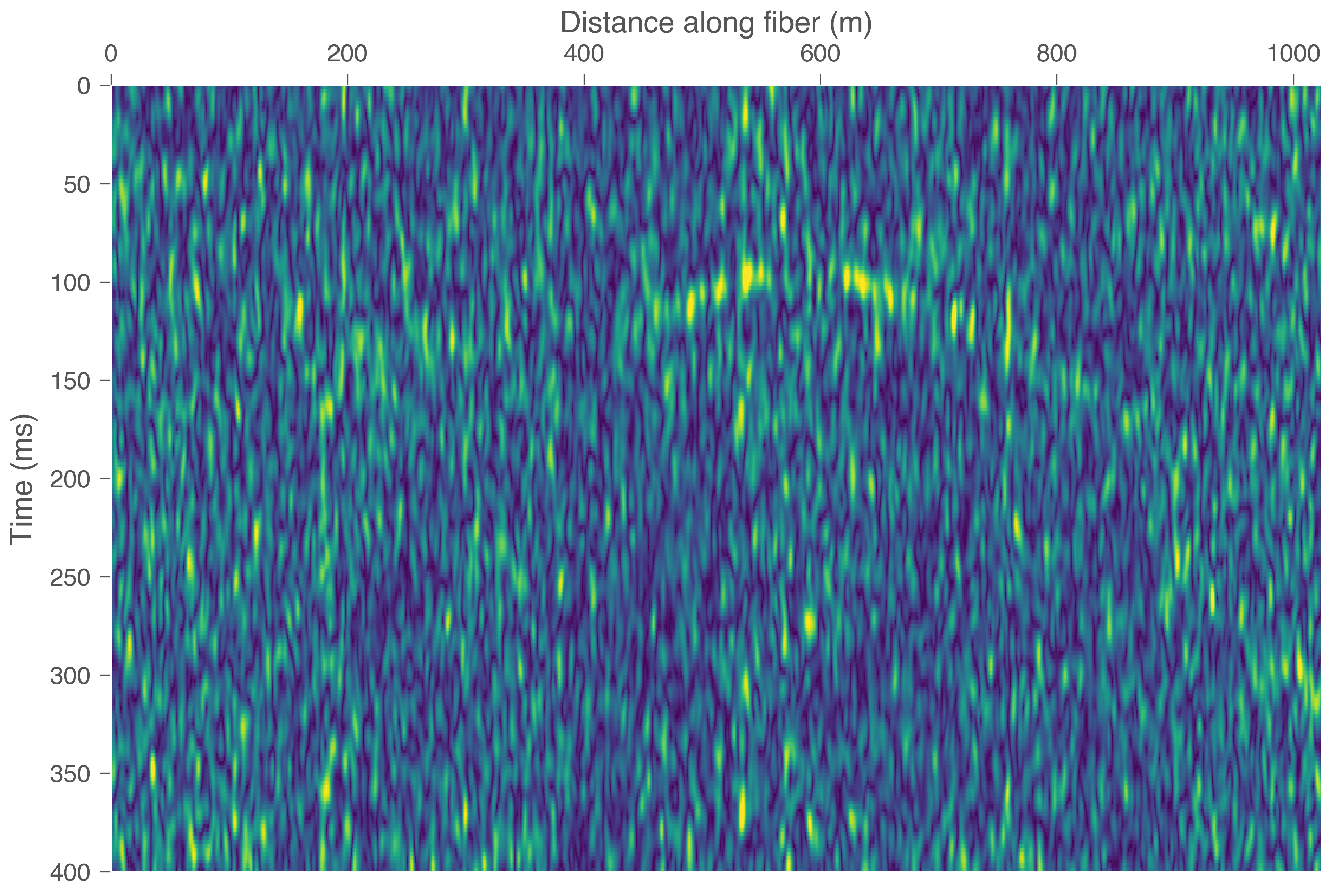}
    \end{center}
    \caption{Two example events that were flagged as microseismic signal by the machine learning model but missed by manual labeling. The seismic sections of these events, on the left, do not show apparent signals. However, the continuous wavelet transform (CWT) analysis of these two sections, on the right, reveals coherent energy over slopes that represent guided wave propagation, confirming that these are indeed microseismic events.}
    \label{false_positives}
\end{figure}

As we examine the metrics over continuous data for this stage (Table \ref{metrics_continuous}), we see that it achieves lower precision than for the test dataset. This discrepancy is an indicator of false positives, but close analysis reveals them as actual events (Figure \ref{false_positives}). These events are too weak to be easily identifiable visually and were not part of the training, evaluation, or test data. A 2-D continuous wavelet transform (CWT) workflow, as described by \cite{huot2019automatic}, shows coherent energy confirming these as actual microseismic events, revealing hundreds of low-amplitude events missed during manual labeling. Consequently, the metrics on the continuous data are not entirely representative of the network's performance as it picks uncatalogued weak events correctly. 

\begin{table}
    \renewcommand{\arraystretch}{1.4}
    \begin{center}
    \caption{Classification metrics for microseismic event detection computed over continuous data of a stimulation stage on which we manually picked all the events.}
    \label{metrics_continuous}
    \begin{tabular*}{\columnwidth}{c @{\extracolsep{\fill}} ccccc}
                    & \bf Accuracy & \bf AUC\textsuperscript{*} & \bf Precision & \bf Recall \\
    \hline
    \hline
    Continuous data & 91.8\%\textsuperscript{$\dagger$} & 93.1\%\textsuperscript{$\dagger$} &  83.3\%\textsuperscript{$\dagger$} & 99.4\% \\
    \end{tabular*}
    \end{center}
    \footnotesize{\textsuperscript{*}Area under the curve. \\
    \textsuperscript{$\dagger$}The marked values are actually not representative of the model's performance. After analysis, many of the events marked as false positives turned out to be actual events missed during manual labeling.}
\end{table}

This analysis suggests that continuous wavelet decomposition could be a candidate feature space for event classification on seismic data. 2-D CWT is a convolution of the input data with rotated, compressed, and dilated analyzing wavelets. Manually labeling the data after applying CWT is challenging because the output of CWT is 5-D. We could train a machine learning model to find the suitable detection thresholds for each frequency and ray parameter of the analyzing wavelet, to capture the varying focal mechanisms of the microseismic events. However, shared weights and pooling layers make CNNs more memory-efficient than CWT. Computational efficiency is desirable because fracture characterization has the most value when conducted in time to influence engineering decisions. 

\subsection{Fracture characterization over multiple stages}

\begin{figure}
    \begin{center}
    \includegraphics[width=0.6\linewidth]{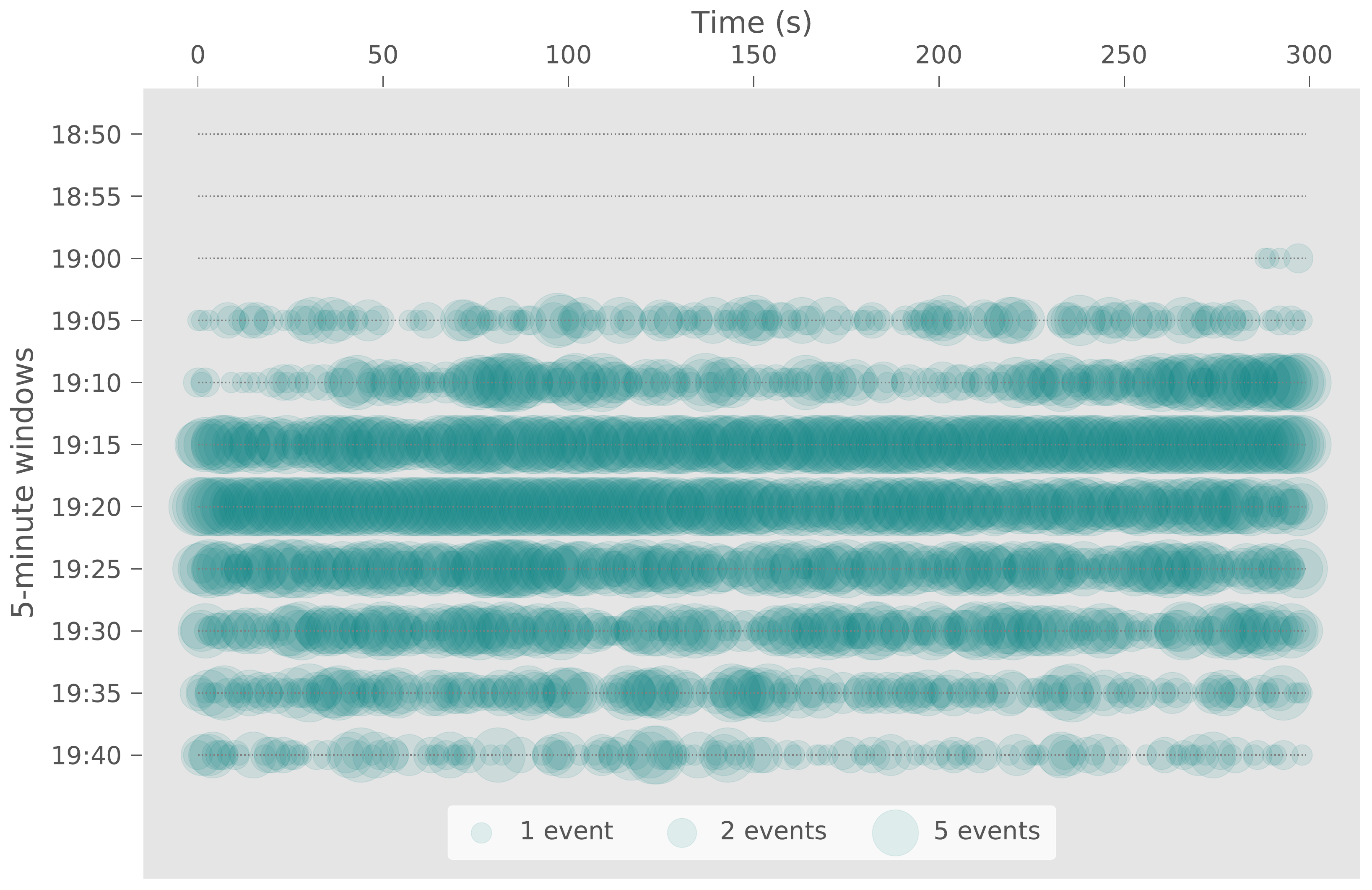}
    \end{center}
    \caption{Microseismic event density detected by our machine learning model for a hydraulic stimulation stage for which there are no hand-picked labels. Each line represents 5 minutes of data, and every circle corresponds to the number of events per second. The model accurately captures the event density pattern ramping up and then going down throughout the stage.}
    \label{predictions}
\end{figure}

\begin{figure}
    \begin{center}
    \includegraphics[width=0.6\linewidth,trim={1.5cm 1.2cm 4.5cm 3cm},clip]{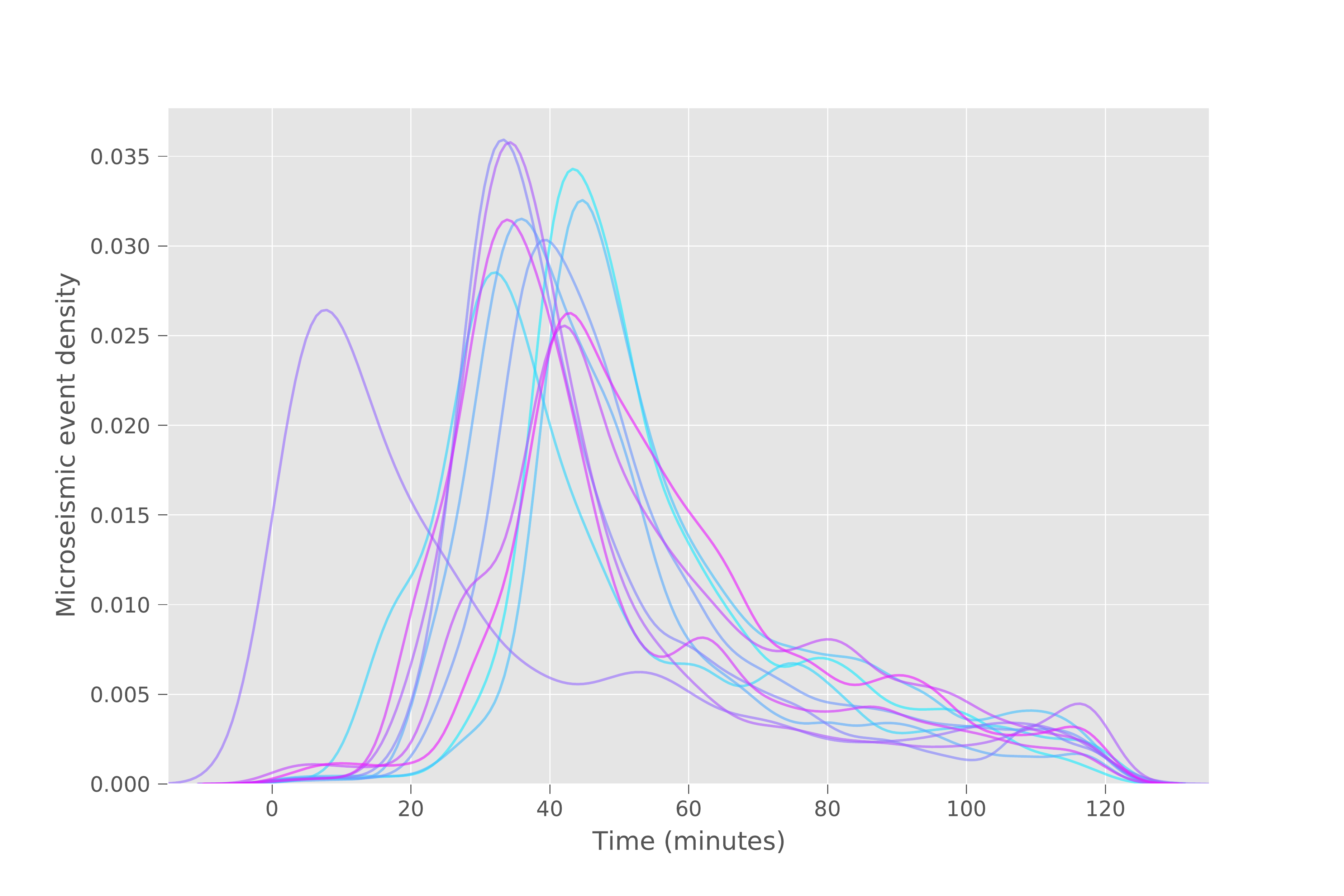}
    \end{center}
    \caption{Microseismic event density detected by our machine learning model over ten hydraulic stimulation stages, each plotted in a different color. Since we cut out the continuous time windows for each stage approximately, the timing of the onset of the peak of the density curve is arbitrary. Nonetheless, the model accurately captures the event density pattern ramping up and then going down throughout each stage.}
    \label{ten_stages}
\end{figure}

We proceed to scale our workflow to multiple stages, as illustrated in Figures~\ref{predictions} and~\ref{ten_stages}. We run the detections over ten different stages, which represent 20 hours of continuous data. The model predicts the event density pattern well and shows the ramping up and slowing down that we expect during each stage. This result indicates good generalization performance from training on two stages to scaling to multiple stages, despite the stages being at different locations with varying, albeit slowly, geology. The results on ten stages yield over 100,000 new microseismic events, demonstrating that this workflow scales to the entire dataset. In contrast, manual picking would be unfeasible to complete in time to influence reservoir stimulation and production decisions (\cite{Verdon2020, Stork2020}).

Given the unidirectional nature of DAS recordings, a single fiber is usually not sufficient to resolve the location of microseismic events. Recent studies (\cite{lellouch2019observations, lellouch2020comparison, lellouch2021properties}) have identified that guided waves can propagate in unconventional reservoirs. As the velocities inside the reservoir are lower than in the surrounding rocks, the reservoir layer traps seismic energy and conducts it for long distances with minimal loss. In particular, \cite{huff2020validating} showed that, for microseismic events, guided waves are generated only if the source is located inside or close to the low-velocity reservoir. All types of guided waves are dispersive, with low frequencies propagating faster than high frequencies. As a result, recorded events displaying a known dispersive nature can be constrained to originating inside or close to the reservoir. In addition, guided waves propagate only inside the reservoir, which means we can approximate their path as purely horizontal. Therefore, by using guided waves, we overcome the cylindrical symmetry problem arising from the directional nature of DAS recordings. We are thus able to obtain microseismic locations for a large portion of the recorded events, despite using DAS recordings from a single horizontal well. 

\begin{figure}
    \begin{center}
    \includegraphics[width=0.7\linewidth,trim={1.5cm 3cm 3.2cm 4.5cm},clip]{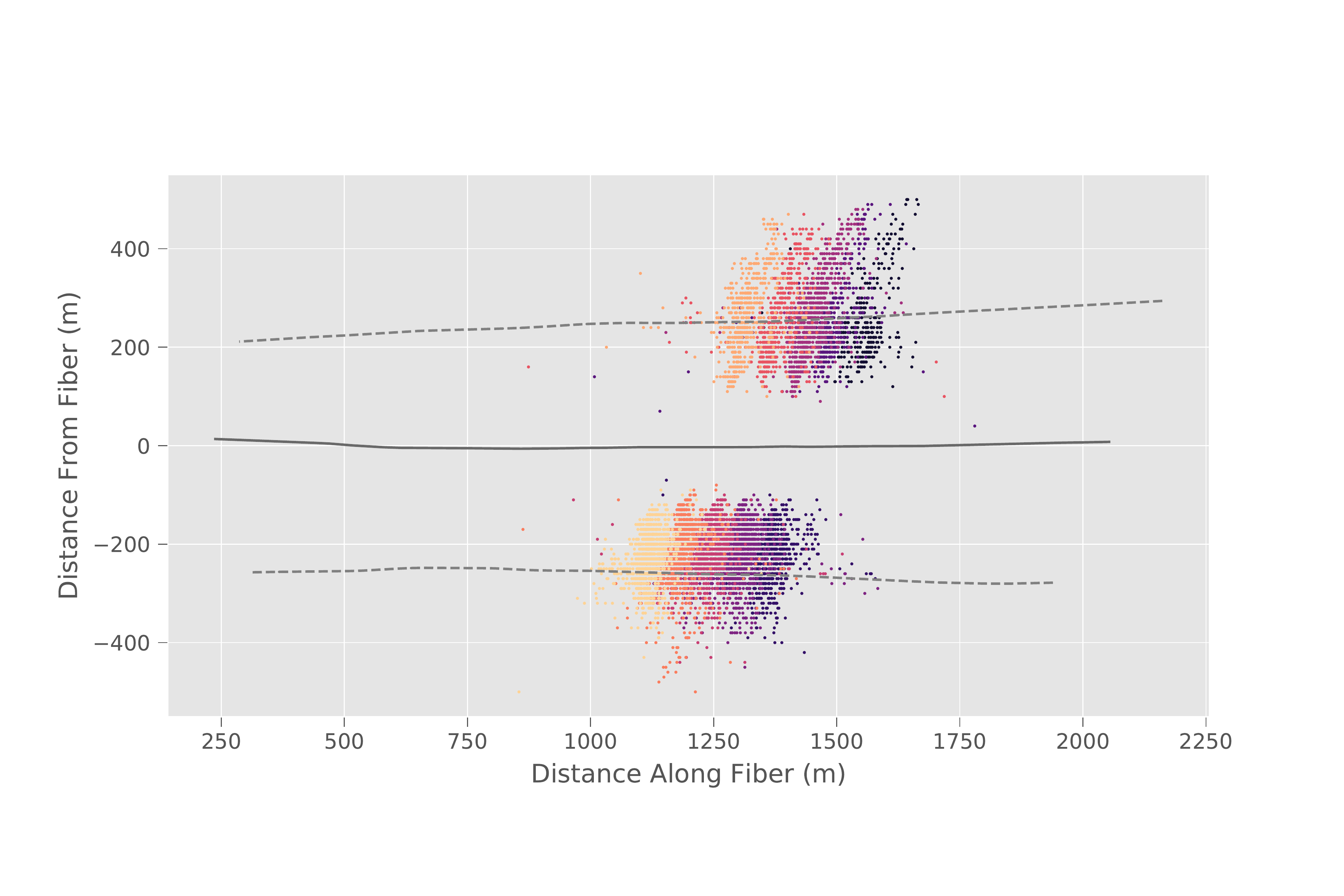}
    \end{center}
    \caption{Microseismic event locations in map view. The middle well, plotted in solid gray, is the one instrumented with fiber. The offset wells are plotted in dotted gray.  Event locations are color-coded based on the stage during which they were recorded. 
    \label{event_locations}}
\end{figure}

We derive the location from the kinematic dispersion properties of the events that originate within the reservoir using the methodology described by \cite{lellouch2022microseismic}. The location in the direction of the fiber is based on an algorithm that automatically estimates the location of the apex of the hyperbolic events. The distance along the orthogonal direction to the fiber is determined by analyzing the frequency dispersion characteristics of the arrival, based on the analysis by \cite{lellouch2021properties}. We apply this location methodology to the 115,270 detected events. We exclude events for which an accurate location can not be computed, thereby automatically excluding any false detections that do not have coherent kinematic properties as well as events that do not originate from the shale reservoir. The results (Figure \ref{event_locations}) reveal a primary propagation direction of fractures in both wells and consistent fracture growth between stages. In the North-East well, fracture growth appears to be towards and away from the fiber. In the South-West well, it is predominantly towards the fiber. The fracture growth direction is consistent across stages and non-perpendicular to the wells, an indicator of regional stress within the subsurface. Further confirmation of this fracture growth direction is provided by the analysis of the perforation shots' arrivals, which show different behaviors for each of the offset wells, as described by \cite{lellouch2021properties}. The high number of detected events allows for this detailed mapping of the fracture locations and propagation directions. This mapping informs us of the geomechanical conditions of the subsurface, primarily the principal stress orientation, and ensures operational safety and efficiency.

\section{Conclusions}
DAS can be installed at low cost in locations where traditional sensors are not easily deployed or operative. This property, combined with its extensive spatial coverage and high resolution, gives it excellent potential for microseismic analysis. We believe that our automated processing workflow opens up multiple paths for widespread DAS-based microseismic analysis for fracture characterization and subsurface monitoring. We demonstrate our methodology on an example dataset from hydraulic stimulation, but as needs for subsurface monitoring technologies grow, it could be a valuable tool for geothermal and CCS applications.

Conventional signal processing approaches, such as trace-by-trace analysis, do not work well for DAS data as the signal-to-noise ratio can vary significantly across traces. In this study, we take full advantage of the two-dimensional spatio-temporal nature of DAS data by using a 2D CNN for the detection task. We detect microseismic events recorded over DAS fiber with accuracy and efficiency much greater than that could be achieved through manual detection. Our machine learning model even detects low signal-to-noise events missed during manual labeling and yields overall better results than other microseismic detections up to date. 

We create a curated catalog of nearly 7,000 manually labeled events to train the machine learning model. This dataset is an order of magnitude larger than previous DAS-based microseismic datasets and covers a range of field data events of different amplitudes and focal mechanisms. In particular, it is large enough to yield good machine learning results without the need to generate synthetic data. In particular, this dataset contains strongly dispersive guided wave events, for which it is challenging to create representative synthetic data without expensive 3-D wave-equation modeling.

We achieve high accuracy by optimizing the network architecture together with its training hyperparameters using a Bayesian optimization scheme. We take heuristics from state-of-the-art classifiers to quickly identify good candidate network architectures. Our resulting neural network achieves excellent performance while having fewer layers. Small networks are easier to train on a limited number of labeled data and less prone to overfitting. Furthermore, they are faster to tune and optimize. This result demonstrates that good performance can be achieved with a quick engineering cycle at low computational cost.

We scale the detection over ten hydraulic stimulation stages and detect over 100,000 new microseismic events. We obtain remarkable generalization performance over multiple stages, even for events located further along the well than the training data examples. From an operational standpoint, this suggests that after initial setup and training, a machine learning model could be used for continuous monitoring without requiring retraining. Additionally, once the model is trained, it can perform detections on continuous data in a matter of seconds, making it valuable for real-time engineering decisions. The location of these events, based on the kinematic dispersion properties of the waves that initiate within the reservoir, allows us to reconstruct a detailed mapping of the propagation direction of the fractures. The estimated fracture growth is consistent across stages, can be reliably interpreted, and is qualitatively validated by perforation shot analysis.

\section{Data and Resources}

The data were provided by Chevron Technical Centre and are proprietary. They cannot be released to the public.
We share our source code for users to adapt to their use cases.

\section{Acknowledgements}
We thank Chevron Technical Centre for making the data available for this research and giving permission to publish this study. We thank the Stanford Exploration Project affiliate members for their financial support.

\newpage



\begin{thebibliography}{}
\itemsep0pt

\bibitem[Abadi et~al., 2015]{tensorflow2015-whitepaper}
Abadi, M., A. Agarwal, P. Barham, E. Brevdo, Z. Chen, C. Citro, G.~S. Corrado,
  A. Davis, J. Dean, M. Devin, S. Ghemawat, I. Goodfellow, A. Harp, G. Irving,
  M. Isard, Y. Jia, R. Jozefowicz, L. Kaiser, M. Kudlur, J. Levenberg, D.
  Man\'{e}, R. Monga, S. Moore, D. Murray, C. Olah, M. Schuster, J. Shlens, B.
  Steiner, I. Sutskever, K. Talwar, P. Tucker, V. Vanhoucke, V. Vasudevan, F.
  Vi\'{e}gas, O. Vinyals, P. Warden, M. Wattenberg, M. Wicke, Y. Yu, and X.
  Zheng (2015).
\newblock {TensorFlow}: Large-scale machine learning on heterogeneous systems.
\newblock (Software available from tensorflow.org).

\bibitem[Baird et~al., 2020]{Baird2020}
Baird, A.~F., A.~L. Stork, S.~A. Horne, G. Naldrett, J.-M. Kendall, J. Wookey,
  J.~P. Verdon, and A. Clarke (2020).
\newblock {Characteristics of microseismic data recorded by distributed
  acoustic sensing systems in anisotropic media}, {\em Geophysics} {\bf 85}
  KS139--KS147.

\bibitem[Bakku, 2015]{bakku2015fracture}
Bakku, S.~K. (2015).
\newblock {\em Fracture characterization from seismic measurements in a
  borehole}, PhD thesis, Massachusetts Institute of Technology.

\bibitem[Binder and Chakraborty, 2019]{Binder2019}
Binder, G., and D. Chakraborty (2019).
\newblock {Detecting microseismic events in downhole distributed acoustic
  sensing data using convolutional neural networks}, in {\em SEG Technical
  Program Expanded Abstracts 2019}, Society of Exploration Geophysicists,
  4864--4868.

\bibitem[Binder and Tura, 2020]{Binder2020}
Binder, G., and A. Tura (2020).
\newblock {Convolutional neural networks for automated microseismic detection
  in downhole distributed acoustic sensing data and comparison to a surface
  geophone array}, {\em Geophys. Prospect.} {\bf 68} 2770--2782.

\bibitem[Daley et~al., 2013]{daley2013field}
Daley, T.~M., B.~M. Freifeld, J. Ajo-Franklin, S. Dou, R. Pevzner, V.
  Shulakova, S. Kashikar, D.~E. Miller, J. Goetz, J. Henninges, et~al. (2013).
\newblock Field testing of fiber-optic distributed acoustic sensing (DAS) for
  subsurface seismic monitoring, {\em The Leading Edge} {\bf 32} 699--706.

\bibitem[Feurer et~al., 2019]{feurer2019auto}
Feurer, M., A. Klein, K. Eggensperger, J.~T. Springenberg, M. Blum, and F.
  Hutter (2019).
\newblock Auto-sklearn: efficient and robust automated machine learning, in
  {\em Automated Machine Learning}, Springer, Cham,  113--134.

\bibitem[Goodfellow et~al., 2016]{goodfellow2016deep}
Goodfellow, I., Y. Bengio, and A. Courville (2016).
\newblock {\em Deep learning}, MIT Press.

\bibitem[He et~al., 2015]{he2015deep}
He, K., X. Zhang, S. Ren, and J. Sun (2015).
\newblock Deep Residual Learning for Image Recognition, {\em arXiv preprint}
  1512.03385.

\bibitem[Huff et~al., 2020]{huff2020validating}
Huff, O., A. Lellouch, B. Luo, G. Jin, and B.~L. Biondi (2020).
\newblock Validating the origin of microseismic events in target reservoir
  using guided waves recorded by DAS, {\em The Leading Edge} {\bf 39} 776--784.

\bibitem[Huot and Biondi, 2018]{huot2018machine}
Huot, F., and B.~L. Biondi (2018).
\newblock Machine learning algorithms for automated seismic ambient noise
  processing applied to DAS acquisition, in {\em SEG Technical Program Expanded
  Abstracts 2018}, Society of Exploration Geophysicists,  5501--5505.

\bibitem[Huot and Biondi, 2020]{huot2020detecting}
--------, (2020).
\newblock Detecting earthquakes through telecom fiber using a convolutional
  neural network, in {\em SEG Technical Program Expanded Abstracts 2020},
  Society of Exploration Geophysicists,  3452--3456.

\bibitem[Huot et~al., 2018a]{huot2018jump}
Huot, F., B.~L. Biondi, and G. Beroza (2018a).
\newblock Jump-starting neural network training for seismic problems, in {\em
  SEG Technical Program Expanded Abstracts 2018}, Society of Exploration
  Geophysicists,  2191--2195.

\bibitem[Huot et~al., 2019]{huot2019automatic}
Huot, F., B.~L. Biondi, A. Lichnewsky, and C. Boneti (2019).
\newblock Automatic denoising by 2-D continuous wavelet transform, in {\em SEG
  Technical Program Expanded Abstracts 2019}, Society of Exploration
  Geophysicists,  3944--3948.

\bibitem[Huot et~al., 2021]{huot2021detecting}
Huot, F., A. Lellouch, P. Given, R.~G. Clapp, B.~L. Biondi, T. Nemeth, and K.
  Nihei (2021).
\newblock Detecting microseismic events on DAS fiber with super-human accuracy,
  Presented at the SEG/AAPG/SEPM First International Meeting for Applied
  Geoscience \& Energy, OnePetro.

\bibitem[Huot et~al., 2017]{huot2017automatic}
Huot, F., Y. Ma, R. Cieplicki, E.~R. Martin, and B. Biondi (2017).
\newblock Automatic noise exploration in urban areas, in {\em SEG Technical
  Program Expanded Abstracts 2017}, Society of Exploration Geophysicists,
  5027--5032.

\bibitem[Huot et~al., 2018b]{huot2018automated}
Huot, F., E.~R. Martin, and B. Biondi (2018b).
\newblock Automated ambient noise processing applied to fiber optic seismic
  acquisition (DAS), in {\em SEG Technical Program Expanded Abstracts 2018},
  Society of Exploration Geophysicists,  4688--4692.

\bibitem[IEA, 2019]{iea2019world}
IEA (2019).
\newblock {\em World Energy Outlook 2019}, IEA, Paris.

\bibitem[Ioffe and Szegedy, 2015]{ioffe2015batch}
Ioffe, S., and C. Szegedy (2015).
\newblock Batch normalization: Accelerating deep network training by reducing
  internal covariate shift, in {\em International conference on machine
  learning},  448--456.

\bibitem[Karrenbach et~al., 2019]{Karrenbach2019}
Karrenbach, M., S. Cole, A. Ridge, K. Boone, D. Kahn, J. Rich, K. Silver, and
  D. Langton (2019).
\newblock {Fiber-optic distributed acoustic sensing of microseismicity, strain
  and temperature during hydraulic fracturing}, {\em Geophysics} {\bf 84}
  D11--D23.

\bibitem[Kingma and Ba, 2014]{kingma2014adam}
Kingma, D.~P., and J. Ba (2014).
\newblock Adam: A method for stochastic optimization, {\em arXiv preprint}
  1412.6980.

\bibitem[Lal, 2004]{lal2004soil}
Lal, R. (2004).
\newblock Soil carbon sequestration to mitigate climate change, {\em Geoderma}
  {\bf 123} 1--22.

\bibitem[Lellouch et~al., 2021a]{lellouch2021properties}
Lellouch, A., E. Biondi, B.~L. Biondi, and M.~A. Meadows (2021a).
\newblock {Properties of a Deep Seismic Waveguide Measured With an Optical
  Fiber}, {\em Physical Review Research} {\bf 3} 13164.

\bibitem[Lellouch et~al., 2019]{lellouch2019observations}
Lellouch, A., S. Horne, M.~A. Meadows, S. Farris, T. Nemeth, and B.~L. Biondi
  (2019).
\newblock DAS observations and modeling of perforation-induced guided waves in
  a shale reservoir, {\em The Leading Edge} {\bf 38} 858--864.

\bibitem[Lellouch et~al., 2020a]{lellouch2020comparison}
Lellouch, A., N.~J. Lindsey, W.~L. Ellsworth, and B.~L. Biondi (2020a).
\newblock {Comparison between distributed acoustic sensing and geophones:
  Downhole microseismic monitoring of the FORGE geothermal experiment}, {\em
  Seismological Research Letters} {\bf 91} 3256--3268.

\bibitem[Lellouch et~al., 2020b]{Lellouch2020b}
Lellouch, A., M.~A. Meadows, T. Nemeth, and B.~L. Biondi (2020b).
\newblock {Fracture properties estimation using distributed acoustic sensing
  recording of guided waves in unconventional reservoirs}, {\em Geophysics}
  {\bf 85} M85--M95.

\bibitem[Lellouch et~al., 2021b]{lellouch2021low}
Lellouch, A., R. Schultz, N.~J. Lindsey, B.~L. Biondi, and W.~L. Ellsworth
  (2021b).
\newblock Low-magnitude seismicity with a downhole distributed acoustic sensing
  array—Examples from the FORGE geothermal experiment, {\em Journal of
  Geophysical Research: Solid Earth} {\bf 126} e2020JB020462.

\bibitem[Lellouch et~al., 2022]{lellouch2022microseismic}
Lellouch, L., B. Luo, F. Huot, R.~G.~C. Clapp, P. Given, E. Biondi, T. Nemeth,
  K.~T. Nihei, and B.~L. Biondi (2022).
\newblock Microseismic analysis over a single horizontal DAS fiber using guided
  waves, {\em Geophysics} .

\bibitem[Martin et~al., 2018]{martin2018seismic}
Martin, E.~R., F. Huot, Y. Ma, R. Cieplicki, S. Cole, M. Karrenbach, and B.~L.
  Biondi (2018).
\newblock A seismic shift in scalable acquisition demands new processing:
  Fiber-optic seismic signal retrieval in urban areas with unsupervised
  learning for coherent noise removal, {\em IEEE Signal Processing Magazine}
  {\bf 35} 31--40.

\bibitem[Metz et~al., 2005]{metz2005ipcc}
Metz, B., O. Davidson, H. De~Coninck, M. Loos, and L. Meyer (2005).
\newblock {\em IPCC special report on carbon dioxide capture and storage},
  Cambridge: Cambridge University Press.

\bibitem[Oldenburg et~al., 2009]{oldenburg2009certification}
Oldenburg, C.~M., S.~L. Bryant, and J.-P. Nicot (2009).
\newblock Certification framework based on effective trapping for geologic
  carbon sequestration, {\em International Journal of Greenhouse Gas Control}
  {\bf 3} 444--457.

\bibitem[Ramachandran et~al., 2017]{ramachandran2017searching}
Ramachandran, P., B. Zoph, and Q.~V. Le (2017).
\newblock Searching for activation functions, {\em arXiv preprint}  1710.05941.

\bibitem[Russakovsky et~al., 2015]{ILSVRC15}
Russakovsky, O., J. Deng, H. Su, J. Krause, S. Satheesh, S. Ma, Z. Huang, A.
  Karpathy, A. Khosla, M. Bernstein, A.~C. Berg, and L. Fei-Fei (2015).
\newblock {ImageNet Large Scale Visual Recognition Challenge}, {\em
  International Journal of Computer Vision (IJCV)} {\bf 115} 211--252.

\bibitem[Shere, 2013]{shere2013renewable}
Shere, J. (2013).
\newblock {\em Renewable: the world-changing power of alternative energy},
  Macmillan.

\bibitem[Simonyan and Zisserman, 2015]{simonyan2015deep}
Simonyan, K., and A. Zisserman (2015).
\newblock Very deep convolutional networks for large-scale image recognition,
  {\em arXiv preprint}  1409.1556.

\bibitem[Snoek et~al., 2012]{snoek2012practical}
Snoek, J., H. Larochelle, and R.~P. Adams (2012).
\newblock Practical bayesian optimization of machine learning algorithms, {\em
  Advances in neural information processing systems}  2951--2959.

\bibitem[Stork et~al., 2020]{Stork2020}
Stork, A.~L., A.~F. Baird, S.~A. Horne, G. Naldrett, S. Lapins, J.~M. Kendall,
  J. Wookey, J.~P. Verdon, A. Clarke, and A. Williams (2020).
\newblock {Application of machine learning to microseismic event detection in
  distributed acoustic sensing data}, {\em Geophysics} {\bf 85} KS149--KS160.

\bibitem[{The GPyOpt authors}, 2016]{gpyopt2016}
{The GPyOpt authors} (2016).
\newblock {GPyOpt}: A bayesian optimization framework in Python.

\bibitem[Verdon et~al., 2020]{Verdon2020}
Verdon, J.~P., S.~A. Horne, A. Clarke, A.~L. Stork, A.~F. Baird, and J.-M.
  Kendall (2020).
\newblock {Microseismic monitoring using a fibre-optic distributed acoustic
  sensor (DAS) array}, {\em Geophysics} {\bf 85} KS89--KS99.

\bibitem[Wilson and Gerard, 2007]{wilson2007carbon}
Wilson, E., and D. Gerard (2007).
\newblock {\em Carbon capture and sequestration: integrating technology,
  monitoring, regulation}, Blackwell Publishing, Ames, IA (United States).

\end{thebibliography}

\newpage

\end{document}